\documentclass{article}

\usepackage{amssymb}
\usepackage{amsfonts}
\usepackage{amsmath}
\usepackage{geometry}
\usepackage[dvips]{epsfig}

\input{tcilatex}

\begin{document}

\title{\textbf{Optimized Negative Dimensional Integration Method (NDIM) and
multiloop Feynman diagram calculation}}
\author{Iv\'{a}n Gonz\'{a}lez\thanks{%
e-mail: ivan.gonzalez@usm.cl} and Iv\'{a}n Schmidt\thanks{%
e-mail: ivan.schmidt@usm.cl} \\
Department of Physics and Center of Subatomic Studies\\
Universidad T\'{e}cnica Federico Santa Mar\'{\i}a \\
Casilla 110-V, Valpara\'{\i}so, Chile}
\maketitle

\begin{abstract}
We present an improved form of the integration technique known as NDIM
(Negative Dimensional Integration Method), which is a powerful tool in the
analytical evaluation of Feynman diagrams. Using this technique we study a $%
\phi ^{3}\oplus \phi ^{4}$ theory in $D=4-2\epsilon $ dimensions,
considering generic topologies of $L$ loops and $E$ independent external
momenta, and where the propagator powers are arbitrary. The method
transforms the Schwinger parametric integral associated to the diagram into
a multiple series expansion, whose main characteristic is that the argument
contains several Kronecker deltas which appear naturally in the application
of the method, and which we call diagram presolution. The optimization we
present here consists in a procedure that minimizes the series multiplicity,
through appropriate factorizations in the multinomials that appear in the
parametric integral, and which maximizes the number of Kronecker deltas that
are generated in the process. The solutions are presented in terms of
generalized hypergeometric functions, obtained once the Kronecker deltas
have been used in the series. Although the technique is general, we apply it
to cases in which there are 2$\;$or$\;$3 different energy scales (masses or
kinematic variables associated to the external momenta), obtaining solutions
in terms of a finite sum of generalized hypergeometric series de 1 and 2
variables respectively, each of them expressible as ratios between the
different energy scales that characterize the topology. The main result is a
method capable of solving Feynman integrals, expressing the solutions as
hypergeometric series of multiplicity $\left( n-1\right) $, where $n$ is the
number of energy scales present in the diagram.
\end{abstract}

\bigskip\bigskip


\bigskip\bigskip


\vfill\newpage

\section{Introduction}

\qquad In Quantum Field Theory the permanent contrast between experimental
measurements and theoretical models has been possible due to the development
of novel and powerful analytical and numerical techniques in perturbative
calculations. The fundamental problem that arises in perturbation theory is
the actual calculation of the loop integrals associated to the Feynman
diagrams, whose solution is specially difficult since these integrals
contain in general both ultraviolet (UV) and infrared (IR) divergences.
Using the dimensional regularization scheme, which extends the
dimensionality of space-time by adding a fractional piece $(D=4-2\epsilon )$%
, it is possible to know the behavior of such divergences in terms of
Laurent expansions with respect to the dimensional regulator $\epsilon $
when it tends to zero. On the other hand, the structure of the integral
associated to the diagram gets increasingly more complicated when the number
of external lines, loops or energy scales is increased, and therefore
finding an analytical solution is extremely difficult. Of the many different
techniques \cite{VSm} that have been developed in order to evaluate
diagrams, we can mention: integration by parts (IBP), contour parametric
integrations in terms of a Mellin-Barnes representation of the diagram, the
differential equations method (DEM), etc.

In the context of the present article, and as comparison, particular mention
has to be made to the Mellin-Barnes integral representation of the diagram,
which is a very powerful technique for finding solutions in cases of
arbitrary propagator powers. These powers can typically occur when tensorial
structures are present, when in the reduction process generate scalar
integrals with different powers of propagators, or when there are
subtopologies associated to massless virtual particles contained in a
propagator, which adds to the corrected propagator a fractional piece
proportional to the dimensional regulator $\epsilon $. Something similar
happens when the Integration by Parts (IBP) technique is used. The
Mellin-Barnes integral representation consists in transforming the momentum
space representation of the diagram into a parameter space representation,
written in terms of multiple contour integrals. As solution of these contour
integrals one obtains generalized hypergeometric series, whose arguments can
be '1' or ratios of the different energy scales that are present in the
diagram, and which give information about the different kinematical regions,
specified in a natural way by the convergence conditions of these functions.
The most interesting aspect of these type of solutions is that they can be
analyzed and expanded in terms of the dimensional regulator $\epsilon $ in
the kinematical region of interest. For this purpose there are known
techniques, algorithms and calculation packages \cite{SMo}, \cite{DMa}, \cite%
{MKa}.

The specific technique that is used here was suggested originally by
Halliday and Ricotta \cite{IRi}, and it is known as NDIM (Negative
Dimensional Integration Method) since it performs an analytical continuation
in the dimension $D$ to negative values. This can be done due to the fact
that the loop integrals have the property of being analytical functions in
arbitrary dimension. In actual practice this technique represents the
diagram in terms of a multiple series whose argument contains a certain
number of Kronecker deltas, and the solutions emerge naturally from
evaluating the sums in the different forms that these Kronecker deltas
allow. The final result can be always expressed in terms of generalized
hypergeometric series. This method has been developed and used by several
authors, among them A. Suzuki, A. Schmidt, E. Santos, C. Anastasiou, E.
Glover and C. Oleari, in many applications to essentially one-loop diagrams.

Since the results obtained in both the Mellin-Barnes representation of
diagrams and in this technique are generalized hypergeometric functions, it
is possible to compare them. In these series solutions the arguments are
ratios of the different energy scales that appear in the diagram, and the
parameters that characterize them are linear combinations of the propagator
powers and the dimension $D$. In such applications it was shown that at
one-loop level \cite{CAn2,ASu11,ASu13,ASu14,CAn1} this technique is
comparable in terms of the complexity of the solutions, with the one coming
from the Mellin-Barnes representation. Nevertheless, in two or more loop
cases \cite{ASu2,ASu3,ASu4,ASu5,ASu6,ASu10}, the conclusion is that the
generalization of the formalism is not adequate since the solutions get to
be complicated in two essential aspects. First, the number of terms or
contributions of the obtained hypergeometric functions is very high, being
for some two-loop diagrams typically of the order of thousands; and second,
the complexity of each of them increases, due to the fact that the
multiplicity of summations of each hypergeometric series is also quite high
and goes over the number needed to represent the solutions with respect to
the number of kinematical variables that the diagram possesses. Both of
these aspects make that any analytical study of the solutions to the $L$%
-loop case be very complex, and therefore it becomes impossible to identify
each of the contributions associated to a specific kinematical region.
Nevertheless, this technique has an important advantage, since it is much
simpler to apply that the Mellin-Barnes representation, because it changes
the process of solving multiple contour integrals into solving a linear
system of equations, a fact which is very important when one is trying to
find analytical solutions.

We have verified that in the form presented by the above mentioned authors,
the application of this method to the general $L$-loop case is indeed very
cumbersome. This can be observed in the $L-linear$ and $(L+1)-linear$
multinomials that are present in the Schwinger parametric integral, which
contains a number of terms that increases rapidly when the number of loops
or independent external momenta increases. Therefore the number of multiple
expansions grows with respect to the number of constraints or Kronecker
deltas which can be obtained from the mathematical structure of the diagram
and which are an essential part of the method. In the present work we have
deduced the method in a different way, and in the process found analytical
solutions to $L-loop$ diagrams which moreover can contain massive
propagators, extending significantly in this manner the number the diagrams
that can be treated with this technique. In our formalism it is not
necessary to work with the momentum representation of the graph. We start
with its Schwinger parametric representation, using for this purpose the
mathematical structure of a generic $L$ loops, $N$ propagators and $E$
independent external lines diagram, shown in Ref. \cite{IGo}, and which
provides a quick and direct way of obtaining the parametric representation
integrand without explicitly solving the momenta integrals. We concluded
that this technique is quite powerful and has significant advantages in the
evaluation of an infinite set of diagrams belonging to certain topological
classes, which is something that will be shown in the specific examples that
are going to be presented in this work.

Our emphasis in this work will be mainly in finding appropriate procedures
for obtaining the general solutions associated to a diagram, instead of
analyzing in more detail each one of them, although we do it in cases in
which they are naturally related to the convergence relations of the
hypergeometric series, which in turn define the kinematical regions where
these series are valid. The presentation of this work is as follows: section
$\left( II\right) $ contains a description of the algebraic components of
NDIM that are required for Feynman diagram evaluation, with emphasis on the
basic formulae of the method, which are related to delta Kronecker
generation, topology related series expansions, and the reasons for naming
this series multiregion expansion. We also explain here how to obtain in a
simple way the explicit Schwinger parametric representation in the different
energy scales of a general topology \cite{IGo}. Finally we describe an
algorithm that sums up the above in a systematic sequence of steps for
solving Feynman%
\'{}%
s integrals and thus obtaining the multiregion expansion of the diagram, or
which could be called presolution of the graph. In section $\left(
III\right) $ we have considered three examples where the technique is
applied systematically, starting from a simple known case until a complex
one with until now unknown solutions. The first is the Bubble diagram $%
\left( E=1,N=2,L=1\right) $, with equal mass $\left( m\right) $ propagators.
This is a basic example, whose solutions are well known \cite{EBo,ADa1,ADa2}%
. The following example corresponds to the diagram called CBox $\left(
E=3,N=5,L=2\right) $, which although has known solutions \cite{COl}, in the
on-shell massless case this is the first time in which they have been
obtained using this method. Increasing the complexity, we solve this diagram
for the case of three massive propagators, which by itself is a new result.
Finally we find the solution for a four-loop propagator case $\left(
E=1,N=8,L=4\right) $, where the idea is to show that this technique is very
efficient and effective when one factorizes and expands systematically the
multinomials of the parametric representation. Initially the massless
propagator case is solved, and then with two massive propagators. The
massless case is quite simple, since it is a diagram which is reducible loop
by loop; but the second case is much more difficult, and the solution is in
terms of hypergeometric series of the type $_{9}F_{8}$, in the variables $%
\left( p^{2}/4m^{2}\right) $ or its reciprocal $\left( 4m^{2}/p^{2}\right) $%
, according to the kinematical region of interest. This result is relevant
given the difficulty of obtaining it using any other method in the case of
arbitrary propagator powers. In section $(IV)$ we discuss some conceptual
aspects of the integration method and about the complexity of the solutions
when the number of loops $L$ increases. We have added a section $\left(
V\right) $ as an extension of $\left( III\right) $, in which we present
several topologies that contain two or three different energy scales, and
where we only show the presolution in each case. Finally in the appendices
we include a summary of definitions and properties of one and two variable
hypergeometric functions, which appear frequently in the solutions of
Feynman integrals; and also a more detailed description of the equations
that support the integration method here presented, with some of its
properties.

The most important aspect of this work is that we have been able to
associate this technique with a certain family of $L$-loops topologies, with
respect to which it is applied very easily and directly, with the
possibility of solving a large number of Feynman diagrams, which can even
contain massive propagators with arbitrary powers. This allows to extend
considerably the known solutions of Feynman diagrams.

\section{Mathematical foundations}

\qquad In order to understand the method, we describe now its main
components:

\begin{description}
\item[-] Obtaining Schwinger's parametric representation..

\item[-] Foundations of the integration method NDIM.

\item[-] General algorithm.
\end{description}

\subsection{The parametric representation}

\qquad Let us consider a generic topology $G$ that represents a Feynman
diagram in a scalar theory, and suppose that this graph is composed of $N$
propagators or internal lines, $L$ loops (associated to independent internal
momenta $\underline{q}=\left\{ q_{1},...,q_{L}\right\} $, and $E$
independent external momenta $\underline{p}=\left\{ p_{1},...,p_{E}\right\} $%
. Each propagator or internal line is characterized by an arbitrary and in
general different mass, $\underline{m}=\left\{ m_{1},...,m_{N}\right\} $.

Using the prescription of dimensional regularization we can write the
momentum integral expression that represents the diagram in $D=4-2\epsilon $
dimensional Minkowski space as:

\begin{equation}
G=G(\underline{p},\underline{m})=\dint \frac{d^{D}q_{1}}{i\pi ^{\frac{D}{2}}}%
...\frac{d^{D}q_{L}}{i\pi ^{\frac{D}{2}}}\frac{1}{(B_{1}^{2}-m_{1}^{2}+i0)^{%
\nu _{1}}}...\frac{1}{(B_{N}^{2}-m_{N}^{2}+i0)^{\nu _{N}}}.  \label{f11}
\end{equation}%
In this expression the symbol $B_{j}$ represents the momentum of the $j$
propagator or internal line, which in general depends on a linear
combination of external $\left\{ \underline{p}\right\} $ and internal $%
\left\{ \underline{q}\right\} $ momenta: $B_{j}=B_{j}(\underline{q},%
\underline{p})$. We also define $\underline{\nu }=\left\{ \nu _{1},...,\nu
_{N}\right\} $ as the set of powers of the propagators, which in general can
take arbitrary values. After introducing Schwinger's representation, it is
possible to solve the momenta integrals in terms of Gaussian integrals. The
result is Schwinger's parametric representation of the diagram associated to
equation $\left( \ref{f11}\right) $, which in the general case is given by
the expression:

\begin{equation}
G=\dfrac{(-1)^{-\frac{LD}{2}}}{\tprod\nolimits_{j=1}^{N}\Gamma (\nu _{j})}%
\dint\limits_{0}^{\infty }d\overrightarrow{x}\;\dfrac{\exp \left(
\tsum\nolimits_{j=1}^{N}x_{j}m_{j}^{2}\right) \exp \left( -\dfrac{F}{U}%
\right) }{U^{\frac{D}{2}}}.  \label{f12}
\end{equation}%
We have introduced for simplicity the following notation $d\overrightarrow{x}%
=dx_{1}...dx_{N}\,\tprod\nolimits_{j=1}^{N}x_{j}^{\nu _{j}-1}$, where $U$
and $F$ are $L-linear$ and $(L+1)-linear$ multinomials respectively, defined
in terms of determinants \cite{IGo}:

\begin{equation}
\begin{array}{ll}
U & =\Delta _{LL}^{(L)}, \\
&  \\
F & =\dsum\limits_{i,j=1}^{E}\Delta _{(L+i)(L+j)}^{(L+1)}\;p_{i}.p_{j} \\
&  \\
& =\dsum\limits_{i=1}^{E}\Delta
_{(L+i)(L+j)}^{(L+1)}\;p_{i}^{2}+2\dsum\limits_{i=1}^{E-1}\dsum%
\limits_{j>i}^{E}\Delta _{(L+i)(L+j)}^{(L+1)}\;p_{i}.p_{j},%
\end{array}%
\end{equation}%
where $\Delta _{ij}^{(k+1)}$ is the determinant defined by the equation:

\begin{equation}
\Delta _{ij}^{(k+1)}=\left\vert
\begin{array}{cccc}
M_{11} & \cdots & M_{1k} & M_{1j} \\
\vdots &  & \vdots & \vdots \\
M_{k1} & \cdots & M_{kk} & M_{kj} \\
M_{i1} & \cdots & M_{ik} & M_{ij}%
\end{array}%
\right\vert .
\end{equation}%
The matrix symmetric $\mathbf{M}$ has dimension $\left( L+E\right) \times
\left( L+E\right) $, and is called Initial Parameters Matrix. It is possible
to easily built it when $\left( \ref{f11}\right) $ is parametrized, and the
internal products of loop momenta and external momenta are expanded, with
coefficients which correspond to the elements of the matrix $M_{ij}$. For a
better understanding of this process let us define the momentum:%
\begin{equation}
Q_{j}=\left\{
\begin{array}{lll}
q_{j} & \text{if} & L\geq j\geq 1, \\
&  &  \\
p_{j-L} & \text{if} & \left( L+E\right) \geq j>L,%
\end{array}%
\right.
\end{equation}%
with the $\left( L+E\right) $-$vector$ $\mathbf{Q=[}Q_{1}\;Q_{2}\;...%
\;Q_{(L+E)}]^{t}$. Using this definition the following matrix structure is
generated, as a previous step to the loop momenta integration:%
\begin{equation}
\dsum\limits_{i=1}^{L+E}\dsum\limits_{j=1}^{L+E}Q_{i}M_{ij}Q_{j}=\mathbf{Q}%
^{t}\mathbf{MQ},
\end{equation}%
with which we finally identify the symmetric matrix $\mathbf{M}$. Let us
develop briefly a simple example, for the massless one loop diagram shown in
Fig. 1.

\begin{figure}[th]
\begin{center}
\epsfig{file=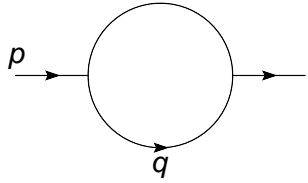,width=0.25\textwidth}
\end{center}
\caption{Bubble diagram.}
\end{figure}
The corresponding 1-loop integral is given by the expression:

\begin{equation}
G=\int \frac{d^{D}q}{i\pi ^{D/2}}\frac{1}{q^{2}(q+p)^{2}}.
\end{equation}%
We then apply Schwinger's parametrization, and then obtain the following
equation:

\begin{equation}
G=\int \frac{d^{D}q}{i\pi ^{D/2}}\int\limits_{0}^{\infty }dx_{1}\;\exp
\left( -x_{1}\;q^{2}\right) \int\limits_{0}^{\infty }dx_{2}\;\exp \left(
-x_{2}\left( q+p\right) ^{2}\right) ,
\end{equation}%
or equivalently, expanding the squares in the exponents and factorizing in
terms of internal and external momenta, it follows that:

\begin{equation}
G=\int dx_{1}dx_{2}\int \frac{d^{D}q}{i\pi ^{D/2}}\;\exp \left[ -\left[
\left( x_{1}+x_{2}\right) \;q^{2}+2x_{2}\;q.p+x_{2}\;p^{2}\right] \right] .
\end{equation}%
Now the matrix of parameters is obtained directly from the exponent by
simple observation, that is:

\begin{equation}
\mathbf{M}=\left(
\begin{array}{cc}
x_{1}+x_{2} & x_{2} \\
x_{2} & x_{2}%
\end{array}%
\right) ,
\end{equation}%
where it has been used that $Q_{1}=q$ and $Q_{2}=p$. Later on other examples
are going to be presented. For more details see Ref. \cite{IGo}.

\subsection{Foundations of the integration method NDIM}

\subsubsection{Notation and essential formulae}

\qquad The exponential structure that the integral of parameters $\left( \ref%
{f12}\right) $ presents allows that the integration method NDIM be developed
and sustained by two formulae that can be deduced starting from the
following integral expression:

\begin{equation}
\frac{1}{A^{\beta }}=\frac{1}{\Gamma (\beta )}\dint\limits_{0}^{\infty
}dx\;x^{\beta -1}\exp (-Ax),  \label{f20}
\end{equation}%
where the quantities $A$ and $\beta $ are in general complex. Using this
equality one can justify the essential point of this method, and which is
related to the equivalence (from an operational point of view) of the
integration sign with a quantity which is proportional to a Kronecker delta.
In order to show this we expand the integrand of equation $\left( \ref{f20}%
\right) $:

\begin{equation}
\frac{1}{A^{\beta }}=\frac{1}{\Gamma (\beta )}\dsum\limits_{n}\frac{\left(
-1\right) ^{n}}{\Gamma \left( n+1\right) }A^{n}\dint\limits_{0}^{\infty
}dx\;x^{\beta +n-1}.
\end{equation}%
Now the corresponding evaluation of the integral will not be done in the
usual way, but we define an identity which fulfills the equality. This
happens after we define the following operational relation:

\begin{equation}
\dint\limits_{0}^{\infty }dx\;x^{\beta +n-1}\equiv \Gamma \left( \beta
\right) \dfrac{\Gamma \left( n+1\right) }{\left( -1\right) ^{n}}\;\delta
_{\beta +n,0}.  \label{f28}
\end{equation}%
For convenience we introduce the following notation in order to write the
operational equivalent of the integral sign:

\begin{equation}
\dint dx\;x^{\alpha +\beta -1}\equiv \left\langle \alpha +\beta
\right\rangle ,  \label{f21}
\end{equation}%
where the parenthesis $\left\langle \cdot \right\rangle $ has implicit the
constraint associated to the Kronecker delta and which furthermore satisfies
the following property (see appendix):

\begin{equation}
\left\langle \alpha +\beta \right\rangle \equiv \Gamma (-\omega )\dfrac{%
\Gamma (\omega +1)}{(-1)^{\omega }}\;\delta _{\alpha +\beta ,0},
\end{equation}%
or in a simplified manner:

\begin{equation}
\left\langle \alpha +\beta \right\rangle =\frac{\Gamma (-\omega )}{\phi
_{\omega }}\;\delta _{\alpha +\beta ,0},  \label{f35}
\end{equation}%
where $\omega $ is an arbitrary index or parameter and where we have also
defined the factor:

\begin{equation}
\phi _{\omega }=\dfrac{(-1)^{\omega }}{\Gamma (\omega +1)}.  \label{f38}
\end{equation}%
On the other hand, by applying successively equation $\left( \ref{f20}%
\right) $ and then equation $\left( \ref{f21}\right) $ to an arbitrary
multinomial, we can deduce the second fundamental formula on which the
method is based. This expresses the fact that a multinomial of $\sigma $
terms can be written as a multiregion expansion, in such a way that it
simultaneously contains all limiting possible cases with respect to the
relation between the different terms present in the multinomial. The series
so described can be written as:

\begin{equation}
\left( A_{1}+...+A_{\sigma }\right) ^{\pm \nu
}=\dsum\limits_{n_{1}}...\dsum\limits_{n_{\sigma }}\phi _{n_{1},..,n_{\sigma
}}\ A_{1}^{n_{1}}...A_{\sigma }^{n_{\sigma }}\frac{\left\langle \mp \nu
+n_{1}+...+n_{\sigma }\right\rangle }{\Gamma (\mp \nu )},  \label{f22}
\end{equation}%
where the terms $A_{i}$ $\left( i=1,...,\sigma \right) $ and the exponent $%
\nu $ are quantities that can take arbitrary values. The expression $\left( %
\ref{f22}\right) $ contains the $\sigma $ expansions that can be obtained
and which have the general form:

\begin{equation}
\sim \frac{A_{i}^{\pm \nu }}{\Gamma (\mp \nu )}\dsum\limits_{n_{1}=0}^{%
\infty }...\dsum\limits_{n_{\sigma }=0}^{\infty }\phi _{n_{1},..,n_{\sigma
}}\;\left( \frac{A_{1}}{A_{i}}\right) ^{n_{1}}...\left( \frac{A_{\sigma }}{%
A_{i}}\right) ^{n_{\sigma }}\Gamma \left( \tsum\limits_{j=1}^{\sigma
}n_{j}\mp \nu \right) \;\delta _{n_{i},0},  \label{f26}
\end{equation}%
where the definition of the factor in $\left( \ref{f38}\right) $ has been
generalized:
\begin{equation}
\phi _{n_{1},..,n_{\sigma }}=(-1)^{_{n_{1}+...+n_{\sigma }}}\dfrac{1}{\Gamma
(n_{1}+1)...\Gamma (n_{\sigma }+1)}.
\end{equation}%
In this case each of the expansions $\left( \ref{f26}\right) $ that can be
obtained starting from $\left( \ref{f22}\right) $ correspond to
multivariable generalized hypergeometric series of multiplicity $\mu =\left(
\sigma -1\right) $, and each one of them contains the limiting cases or
region where it is fulfilled that $\left( \dfrac{A_{j}}{A_{i}}<1\right) $ $%
\forall $ $j\neq i$.

\subsubsection{General form of the multiregion expansion (MRE) of a diagram
and its solution}

\qquad Once the parametric representation of a diagram $\left( \ref{f12}%
\right) $ has been obtained, the following step is to evaluate it, and for
this purpose it is necessary to make an expansion of the integrand starting
from the exponential functions if they are present, and then continuing with
a multiregion expansion of all the multinomials that the process generates
according to formula $\left( \ref{f22}\right) .$ The expansion process ends
when one finally obtains only one term, which is a product of all the
Schwinger parameters. Now it only remains to replace all the integrals
according to the formula $\left( \ref{f21}\right) $\ or its equivalent $%
\left\langle \cdot \right\rangle $. We have obtained in this manner the
presolution or multiregion expansion of the Feynman integral considered in $%
\left( \ref{f11}\right) $. In the case in which we consider a generic
topology $G$, characterized by $M$ different mass scales, $N$ propagators
and $P$ Lorentz invariants associated to internal products of the external
independent momenta, then the general form of the multiregion expansion is
given by the following expression:

\begin{equation}
G=(-1)^{-\frac{LD}{2}}\dsum\limits_{n_{1},..,n_{\sigma }}\phi
_{n_{1},..,n_{\sigma
}}\;\tprod\limits_{j=1}^{P}(Q_{j}^{2})^{n_{j}}\tprod%
\limits_{j=P+1}^{P+M}(-m_{j}^{2})^{n_{j}}\tprod\limits_{j=1}^{N}\dfrac{%
\left\langle \nu _{j}+\alpha _{j}\right\rangle }{\Gamma (\nu _{j})}%
\tprod\limits_{j=1}^{K}\dfrac{\left\langle \beta _{j}+\gamma
_{j}\right\rangle }{\Gamma (\beta _{j})},  \label{f23}
\end{equation}%
where it is possible to identify the following quantities:

$\sigma \Longrightarrow $ Corresponds to the multiplicity or number of
summations that conform the multiregion expansion, or in this case the
presolution of the diagram $G$.

\bigskip

$Q_{j}^{2}\Longrightarrow $ Corresponds to a kinematical Lorentz invariant,
which is a quadratic form of the independent external momenta.

\bigskip

$\alpha _{j},\beta _{j},\gamma _{j}\Longrightarrow $ Correspond to linear
combinations of the indexes $\left\{ n_{1},..,n_{\sigma }\right\} $, with
the exception of $\beta _{1}$, which includes a dependence on the dimension $%
D$:

\begin{equation}
\beta _{1}=\frac{D}{2}+n_{1}+...+n_{P}.
\end{equation}%
The coefficients of the sum indexes $\left\{ n_{i}\right\} $ in the linear
combinations $\alpha _{j}$ and $\gamma _{j}$ are (+1) and in the case of $%
\beta _{j}$ the indexes have coefficients (-1), except for $\beta _{1}$.

\bigskip

$N\Longrightarrow $Number of propagators or equivalently number of
parametric integrations that the method transforms into $N$ Kronecker deltas.

\bigskip

$K\Longrightarrow $Number related to the total number of multiregion
expansions performed over the integrand of the parametric representation,
which in turn generates $K$ constraints or equivalently $K$ Kronecker deltas.

\bigskip

A particular case corresponds to vacuum fluctuation diagrams, in which case
the presolution $\left( \ref{f23}\right) $ is written in the following form:

\begin{equation}
G=(-1)^{-\frac{LD}{2}}\dsum\limits_{n_{1},..,n_{\sigma }}\phi
_{n_{1},..,n_{\sigma
}}\;\tprod\limits_{j=1}^{M}(-m_{j}^{2})^{n_{j}}\tprod\limits_{j=1}^{N}\frac{%
\left\langle \nu _{j}+\alpha _{j}\right\rangle }{\Gamma (\nu _{j})}%
\tprod\limits_{j=1}^{K}\dfrac{\left\langle \beta _{j}+\gamma
_{j}\right\rangle }{\Gamma (\beta _{j})},
\end{equation}%
where $\beta _{1}=\frac{D}{2}$.

In order to find the solutions, it is necessary to evaluate the sums using
for this purpose the existing constraints among the indexes of the sum,
represented by the $\delta =(N+K)$ Kronecker deltas; but as it can be seen
there are several ways to do this evaluation. In fact the number of
different forms to evaluate the presolution of $G$ using the Kronecker
deltas is given by the combinatorial formula:

\begin{equation}
C_{\delta }^{\sigma }=\dfrac{\sigma !}{\delta !(\sigma -\delta )!}.
\end{equation}%
Each one of these forms of summing using the Kronecker deltas in $G$,
generates in turn as a result a multiple series, which corresponds to a
generalized hypergeometric function, whose multiplicity is given by:

\begin{equation}
\mu =\left( \sigma -\delta \right) .
\end{equation}%
In general it is not always possible to use the $\delta $\ deltas in order
to evaluate the corresponding $\delta $ sums, since this will depend on the
combination of indexes with respect to which the sum should be done. If this
happens, these cases are simply not considered as contributions to the
solution and therefore are discarded. In those cases in which we do have a
relevant expansion contribution, this corresponds to an serie representation
of the set of kinematical variables that are present in the problem, in the
form of ratios of the different energy scales that appear in the topology.
Since the multiregion expansion $\left( \ref{f23}\right) $ contains all the
limits simultaneously, all the solutions that have been found are related
between them by analytical continuation, which is realized implicitly by the
integration method NDIM.

In practical terms the idea of the method is to generate finally an
expansion that represents the diagram $G$, the Multiregion Expansion,
characterized by multiplicity $\sigma $, in combination with a number $%
\delta $ of Kronecker deltas. Using this it is possible to find the
solutions in terms of generalized hypergeometric functions of multiplicity $%
\mu $. In this work we study in detail only the topologies where the $\mu $
variables of the different solutions correspond to ratios of the different $%
\left( P+M\right) $ energy scales present in the topology, that is when the
following condition if fulfilled:

\begin{equation}
\mu =\left( P+M-1\right) .  \label{f46}
\end{equation}

\subsubsection{Minimizing expansions and maximizing the number of
constraints (Kronecker deltas)}

\qquad This is an essential point in the improvement of the integration
technique NDIM, since it explains the procedure used to find the optimal
multiregion expansion of the diagram $G$. The idea is quite simple and it is
related to the form in which the expansion of the multinomials $F$ and $U$
of the parametric representation $\left( \ref{f12}\right) $ is done. The
technique consists in the factorization of all the multinomials which are
repeated two or more times in the integrand, without expanding them until
they factorize in one term as product of the previous expansions. In the
same way the process is repeated for all the submultinomials that the
process generates and ends when finally one obtains one single monomial
which is a product of the $N$ Schwinger parameters. Finally applying
equation $\left( \ref{f21}\right) $, the integrals get associated to $N$
Kronecker deltas.

The main result that is achieved by such a procedure is the minimization of
the number of expansions, increasing the number of submultinomials that have
to be expanded and since each multiregion expansion done over them generates
one Kronecker delta, using formula $\left( \ref{f22}\right) $ it is clear
that the number of deltas is also maximized.

Let us see the following examples where it is verified that this systematic
expansion indeed generates generalized hypergeometric functions of minimized
multiplicity $\mu $, and that in general less terms than in previous
approaches are part of the solution of the diagram.

\paragraph{Example I}

Let us consider the following function:

\begin{equation}
g=(a_{1}a_{2}+a_{1}a_{3}+a_{2}a_{4}+a_{3}a_{4})^{\beta },
\end{equation}%
and let us find its corresponding multiregion expansion, with and without
the factorization previously indicated.

\subparagraph{Expansion without factorization :}

Applying directly formula $\left( \ref{f22}\right) $ to function $g$, one
easily obtains the multiregion expansion of the function:

\begin{equation*}
g=\dsum\limits_{n_{1},..,n_{4}}\phi
_{n_{1},..,n_{4}}%
\;a_{1}^{n_{1}+n_{2}}a_{2}^{n_{1}+n_{3}}a_{3}^{n_{2}+n_{4}}a_{4}^{n_{3}+n_{4}}%
\frac{\left\langle -\beta +n_{1}+n_{2}+n_{3}+n_{4}\right\rangle }{\Gamma
(-\beta )}.
\end{equation*}

\subparagraph{Expansion with factorization :}

We now reorder the function $g$, factorizing the multinomial in the
following way:

\begin{equation}
g=\left[ a_{1}b+a_{4}b\right] ^{\beta },  \label{f24}
\end{equation}%
where $b=\left( a_{2}+a_{3}\right) $. We have applied the previously stated
idea with respect to repeated submultinomials. We now expand the binomial in
$\left( \ref{f24}\right) $, obtaining the following:

\begin{equation}
g=\dsum\limits_{n_{1},n_{2}}\phi
_{n_{1},n_{2}}\;b^{n_{1}+n_{2}}a_{1}^{n_{1}}a_{4}^{n_{2}}\frac{\left\langle
-\beta +n_{1}+n_{2}\right\rangle }{\Gamma (-\beta )},
\end{equation}%
in an analogous manner the binomial $b=\left( a_{2}+a_{3}\right) $ is now
expanded, which gives as result the multiregion expansion of $g$:

\begin{equation}
g=\dsum\limits_{n_{1},..,n_{4}}\phi
_{n_{1},..,n_{4}}\;a_{1}^{n_{1}}a_{2}^{n_{3}}a_{3}^{n_{4}}a_{4}^{n_{2}}\frac{%
\left\langle -\beta +n_{1}+n_{2}\right\rangle }{\Gamma (-\beta )}\frac{%
\left\langle -n_{1}-n_{2}+n_{3}+n_{4}\right\rangle }{\Gamma (-n_{1}-n_{2})}.
\end{equation}%
The advantage of expanding after using the factorization of repeated
submultinomials can be clearly appreciated in Table I.

\begin{equation}
\begin{tabular}{lll}
\hline
& without factorization & with factorization \\ \hline
Multiplicity multiregion series $\left( \sigma \right) $ &
\multicolumn{1}{c}{4} & \multicolumn{1}{c}{4} \\
Kronecker deltas associated to the expansion $\left( \delta \right) $ &
\multicolumn{1}{c}{1} & \multicolumn{1}{c}{2} \\
Multiplicity of resulting series $\left( {\mu =\sigma -\delta }\right) $ &
\multicolumn{1}{c}{3} & \multicolumn{1}{c}{2} \\
Possible resulting expansions $\left( {C}_{\delta }^{\sigma }\right) $ &
\multicolumn{1}{c}{4} & \multicolumn{1}{c}{6} \\
Relevant resulting expansions & \multicolumn{1}{c}{4} & \multicolumn{1}{c}{4}
\\ \hline
\end{tabular}
\tag{Table I}
\end{equation}%
Notice that the multiplicity $\mu $ of the produced hypergeometric series
has decreased, although the number of relevant contributions, those that
correspond to a limiting case of $g$, has remained the same.

\paragraph{Example II}

Let us consider a second example, a function $g$ which is a product of two
monomials:

\begin{equation}
g=\left( a_{1}+a_{2}\right) ^{\alpha }(a_{1}+a_{2}+a_{3})^{\beta },
\label{f25}
\end{equation}%
and find its respective multiregion expansion with and without factorization.

\subparagraph{Expansion without factorization:}

We expand each multinomial separately, and then after reordering obtain the
following series for $g$:

\begin{equation}
g=\dsum\limits_{n_{1},..,n_{5}}\phi
_{n_{1},..,n_{5}}\;a_{1}^{n_{1}+n_{3}}a_{2}^{n_{2}+n_{4}}a_{3}^{n_{5}}\frac{%
\left\langle -\alpha +n_{1}+n_{2}\right\rangle }{\Gamma (-\alpha )}\frac{%
\left\langle -\beta +n_{3}+n_{4}+n_{5}\right\rangle }{\Gamma (-\beta )}.
\end{equation}

\subparagraph{Expansion with factorization :}

Let us now see what happens if the repeated submultinomials in $\left( \ref%
{f25}\right) $ are factorized as follows:

\begin{equation}
g=b^{\alpha }\left[ b+a_{3}\right] ^{\beta },
\end{equation}%
where $b=\left( a_{1}+a_{2}\right) .$ Then expanding the binomial, one
obtains the series:

\begin{equation}
g=\dsum\limits_{n_{1},n_{2}}\phi _{n_{1},n_{2}}\;b^{\alpha
+n_{1}}a_{3}^{n_{2}}\frac{\left\langle -\beta +n_{1}+n_{2}\right\rangle }{%
\Gamma (-\beta )},
\end{equation}%
and then expanding the factor $b=\left( a_{1}+a_{2}\right) $, which finally
gives us the multiregion expansion for $g$:

\begin{equation}
g=\dsum\limits_{n_{1},..,n_{4}}\phi
_{n_{1},..,n_{4}}\;a_{1}^{n_{3}}a_{2}^{n_{4}}a_{3}^{n_{2}}\frac{\left\langle
-\alpha -n_{1}+n_{3}+n_{4}\right\rangle }{\Gamma (-\alpha -n_{1})}\frac{%
\left\langle -\beta +n_{1}+n_{2}\right\rangle }{\Gamma (-\beta )}.
\end{equation}%
The important fact is that once again the complexity of the resulting series
representations gets reduced, both in the multiplicity $\mu $ and in the
number of relevant contributions, as can be seen in Table II.

\begin{equation}
\begin{tabular}{lll}
\hline
& without factorization & with factorization \\ \hline
Multiplicity of multiregion series $\left( {\sigma }\right) $ &
\multicolumn{1}{c}{5} & \multicolumn{1}{c}{4} \\
Kronecker deltas associated to the expansion $\left( {\delta }\right) $ &
\multicolumn{1}{c}{2} & \multicolumn{1}{c}{2} \\
Multiplicity of resulting series $\left( {\mu =\sigma -\delta }\right) $ &
\multicolumn{1}{c}{3} & \multicolumn{1}{c}{2} \\
Possible resulting expansions $\left( {C}_{\delta }^{\sigma }\right) $ &
\multicolumn{1}{c}{10} & \multicolumn{1}{c}{6} \\
Relevant resulting expansions & \multicolumn{1}{c}{6} & \multicolumn{1}{c}{5}
\\ \hline
\end{tabular}
\tag{Table II}
\end{equation}%
Although we have presented very simple examples, they illustrate the
advantages of first doing the factorization and then expanding
systematically a multinomial. In these examples we verify that the
multiplicity of each hypergeometric function finally obtained decreases when
using the factorization procedure of repeated submultinomials, and also the
number of relevant expansions obtained from the multiregion series is
reduced as well.

For the particular case of Feynman integrals, the equivalent Schwinger
parametric representation is composed of two multinomials, $F$ and $U$, and
the factorization and expansion process shown in the examples is directly
applicable to them.

\subsection{The algorithm}

\qquad Here we present an algorithm for finding the solution of an arbitrary
Feynman diagram:

\begin{enumerate}
\item Find Schwinger%
\'{}%
s parametric representation $\left( \ref{f12}\right)$ of the Feynman diagram
$G$, characterized by $L$ loops, $N$ propagators, $E$ independent external
momenta and $M$ different masses. In general $\left( M\leq N\right) $.

\item Minimize the number of terms of the multinomial $F$, reordering such
that:
\begin{equation}
F=\dsum\limits_{i=1}^{E}\dsum\limits_{j=1}^{E}\Delta
_{(L+i)(L+j)}^{(L+1)}\;p_{i}.p_{j}=\dsum\limits_{j=1}^{P}f_{j}(%
\overrightarrow{x})\;Q_{j}^{2},
\end{equation}%
where the factors $f_{j}(\overrightarrow{x})$ correspond to multinomials
which depend only on Schwinger%
\'{}%
s parameters, and where $Q_{j}^{2}$ is a kinematical invariant which depends
on the independent external momenta. Thus the integrand of the parametric
representation acquires the following structure:

\begin{equation}
\dfrac{\exp \left( y_{1}m_{1}^{2}\right) ...\exp \left(
y_{M}m_{M}^{2}\right) \exp \left( -\dfrac{f_{1}Q_{1}^{2}}{U}\right) ...\exp
\left( -\dfrac{f_{P}Q_{P}^{2}}{U}\right) }{U^{D/2}}.
\end{equation}%
Being $\left\{ y_{1},...,y_{M}\right\} \subseteq \left\{
x_{1},...,x_{N}\right\} $. In principle, assuming that all masses are
different and non-vanishing, $(M+P)$ independent (between them) expansions
are obtained, only associated to these exponential functions.

\item The following step is finding repeated multinomials susceptible to
factorize, both in the functions $f_{j}(\overrightarrow{x})$ as well as in
the multinomial $U$. For the equal masses case, before expanding the
exponential which contains them, it is convenient to factorize the terms
that build a multinomial already existent in the previously factorized
integrand, and then one proceeds to expand such exponential.

\item Expand the multinomials until finally a single product of Schwinger's
parameters is obtained. Each performed expansion will be associated to a
Kronecker delta.

\item After doing all the expansions, finally it remains to replace the
integral signs by its equivalent $\left\langle \cdot \right\rangle $. By
doing that we add $N$ additional Kronecker deltas. The result is the
presolution of the diagram $G$ or its equivalent multiregion expansion $%
\left( \ref{f23}\right) $.

\item In order to find the contribution or serie representation associated
to a particular combination of the $\mu $ free indexes, the most appropriate
procedure is to solve the linear system corresponding to the constraints
among the indexes, obtained from the Kronecker deltas, and assume that such
indexes in these equations correspond to independent free variables. Thus we
obtain a set of solutions for the indexes that are not free, in terms of the
$\mu $ free indexes, of the parameters $\left\{ \nu _{1},...,\nu
_{N}\right\} $ and of the dimension $D$. Each of these combinations
constitute a generalized hypergeometric function, whose series
representation has multiplicity $\mu $. Not all the combinations of free
indexes generate a solution, in which case the associated linear system
simply has no solution or equivalently the Kronecker deltas cannot eliminate
the $\delta $ sums in the non-free indexes, so the quantity $C_{\delta
}^{\sigma }$ is an upper bound with respect to the number of possible
hypergeometric contributions that are present in the solution of the diagram
$G$.

\item Repeat the previous process for the $C_{\delta }^{\sigma }$ forms of
combining the $\mu $ indexes. Thus we obtain at most $C_{\delta }^{\sigma }$
hypergeometric series, which are classified according to kinematical region
of interest. The solution in each kinematical region corresponds to the
algebraic sum of all the contributions that have the same kinematical
argument or variable. The final result is the evaluation of the diagram $G$
in terms of all its series solutions.
\end{enumerate}

\section{Applications}

\qquad In order to show explicitly the integration technique, let us
consider three applications. The first corresponds to the evaluation of the
one-loop massive propagator. Through this simple and known problem we
present the formalism and proposed notation. In the second example, the CBox
diagram, it is already possible to notice the efficiency of an adequate
factorization. In this case the solution is also known and has been already
obtained using a different method to the one proposed here, which is useful
in order to compare the simplicity of the NDIM with respect to other
methods. We also generalize this problem to cases in which the solution has
not been found until now. And the last example is a four-loop diagram which
has associated a very complex integral given the number of terms that are
present in the polynomials $F$ and $U$, and which nevertheless with an
optimal factorization becomes possible to be solved, even considering one
mass scale in the graph. Is here where the integration technique
demonstrates all its power, obtaining a completely new result and showing
the simplicity of the method.

\subsection{Example I : Massive Bubble diagram}

\qquad This is a case in which the solutions are well known \cite%
{EBo,ADa1,ADa2}, and which we will use in order to introduce the integration
method:

\begin{figure}[th]
\begin{center}
\epsfig{file=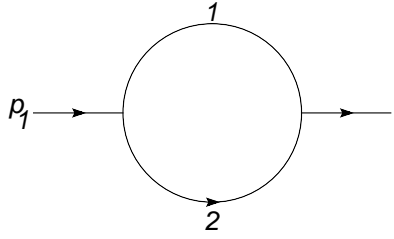,width=0.32\textwidth}
\end{center}
\caption{Labelled Bubble diagram.}
\end{figure}
The integral representation of this diagram (Fig. 2) in momentum space is
given by the equation:

\begin{equation}
G=\int \frac{d^{D}q_{1}}{i\pi ^{D/2}}\frac{1}{(q_{1}^{2}-m_{1}^{2})^{\nu
_{1}}\left( (p_{1}-q_{1})^{2}-m_{2}^{2}\right) ^{\nu _{2}}}.
\end{equation}%
Let us consider the case of equal mass propagators, $m_{1}=m_{2}=m$, and
then according to equation $\left( \ref{f12}\right) $, the corresponding
Schwinger parametric representation is given by the expression:

\begin{equation}
G=\dfrac{(-1)^{-\frac{D}{2}}}{\tprod\nolimits_{j=1}^{2}\Gamma (\nu _{j})}%
\dint\limits_{0}^{\infty }d\overrightarrow{x}\;\frac{\exp \left( -\left(
x_{1}+x_{2}\right) (-m^{2})\right) \exp \left( -\dfrac{x_{1}x_{2}}{%
x_{1}+x_{2}}p_{1}^{2}\right) }{\left( x_{1}+x_{2}\right) ^{\frac{D}{2}}},
\end{equation}%
where we readily identify the multinomials of the representation, that is $%
U=\left( x_{1}+x_{2}\right) $ and $F=x_{1}x_{2}\;p_{1}^{2}$. Then the only
repeated multinomial coincides with $U$. For finding the multiregion
expansion we expand first the exponentials, therefore obtaining the series:

\begin{equation}
G=\dfrac{(-1)^{-\frac{D}{2}}}{\tprod\nolimits_{j=1}^{2}\Gamma (\nu _{j})}%
\dsum\limits_{n_{1},n_{2}}\phi _{n_{1,}n_{2}}\;\left( p_{1}^{2}\right)
^{n_{1}}\left( -m^{2}\right) ^{n_{2}}\dint d\overrightarrow{x}\;\frac{%
x_{1}^{n_{1}}x_{2}^{n_{1}}}{\left( x_{1}+x_{2}\right) ^{\frac{D}{2}%
+n_{1}-n_{2}}}.  \label{k}
\end{equation}%
Now $U$ is expanded:

\begin{equation}
\dfrac{1}{\left( x_{1}+x_{2}\right) ^{\frac{D}{2}+n_{1}-n_{2}}}%
=\dsum\limits_{n_{3},n_{4}}\phi _{n_{3,}n_{4}}\;\frac{%
x_{1}^{n_{3}}x_{2}^{n_{4}}}{\Gamma (\frac{D}{2}+n_{1}-n_{2})}\Delta _{1},
\label{f39}
\end{equation}%
which replaced in $\left( \ref{k}\right) $ and doing a separation of
integration variables gives us:%
\begin{equation}
G=\dfrac{(-1)^{-\frac{D}{2}}}{\tprod\nolimits_{j=1}^{2}\Gamma (\nu _{j})}%
\dsum\limits_{n_{1},..,n_{4}}\phi _{n_{1,..,}n_{4}}\;\frac{\left(
p_{1}^{2}\right) ^{n_{1}}\left( -m^{2}\right) ^{n_{2}}}{\Gamma (\frac{D}{2}%
+n_{1}-n_{2})}\Delta _{1}\int dx_{1}\;x_{1}^{\nu _{1}+n_{1}+n_{3}-1}\int
dx_{2}\;x_{2}^{\nu _{2}+n_{1}+n_{4}-1}.
\end{equation}%
Then using equation $\left( \ref{f21}\right) $ the integrals are transformed
in its equivalent $\left\langle \cdot \right\rangle $, which finally allows
us to obtain the multiregion expansion of the diagram $G$:

\begin{equation}
G=\dfrac{(-1)^{-\frac{D}{2}}}{\tprod\nolimits_{j=1}^{2}\Gamma (\nu _{j})}%
\dsum\limits_{n_{1},..,n_{4}}\phi _{n_{1,..,}n_{4}}\;\left( p_{1}^{2}\right)
^{n_{1}}\left( -m^{2}\right) ^{n_{2}}\frac{\tprod\nolimits_{j=1}^{3}\Delta
_{j}}{\Gamma (\frac{D}{2}+n_{1}-n_{2})},  \label{f27}
\end{equation}%
where we have defined a notation for constraints $\left\{ \Delta
_{i}\right\} $:

\begin{equation}
\begin{array}{l}
\Delta _{1}=\left\langle \frac{D}{2}+n_{1}-n_{2}+n_{3}+n_{4}\right\rangle ,
\\
\Delta _{2}=\left\langle \nu _{1}+n_{1}+n_{3}\right\rangle , \\
\Delta _{3}=\left\langle \nu _{2}+n_{1}+n_{4}\right\rangle .%
\end{array}
\label{f34}
\end{equation}%
The number of possible representations is given by the combinatorics $%
C_{3}^{4}=4$, and the obtained hypergeometric series will have multiplicity
one. The constraints in $\left( \ref{f34}\right) $ provide us with the
following linear system:

\begin{equation}
\begin{array}{l}
0=\frac{D}{2}+n_{1}-n_{2}+n_{3}+n_{4}, \\
\\
0=\nu _{1}+n_{1}+n_{3}, \\
\\
0=\nu _{2}+n_{1}+n_{4}.%
\end{array}
\label{l}
\end{equation}%
Explicitly the combinatorics expresses that we have three equations and four
variables(indexes), and therefore it is necessary to leave one free or
independent index, which can be done in four different ways. Let us see this
case by case:

\subsubsection{Serie representation when $n_{1}$\textit{\ is taken as
independent index:}}

\qquad Before doing any calculation let us define a notation for relating
the index that is taken as free and the respective hypergeometric
representation that it generates. For this purpose let us identify $G_{j}$
as the contribution obtained once we leave index $n_{j}$ free in the
multiregion representation of diagram $G$.

Now starting form the multiregion expression $\left( \ref{f27}\right) $, and
using formula $\left( \ref{f35}\right) $, we can replace the parenthesis $%
\left\langle \cdot \right\rangle $ conveniently, thus obtaining the series
with index of sum $n_{1}$:

\begin{equation}
G_{1}=\dfrac{(-1)^{-\frac{D}{2}}}{\tprod\nolimits_{j=1}^{2}\Gamma (\nu _{j})}%
\dsum\limits_{n_{1}}\dfrac{(-1)^{n_{1}}}{n_{1}!}\left( p_{1}^{2}\right)
^{n_{1}}\left( -m^{2}\right) ^{n_{2}}\frac{\Gamma (-n_{2})\Gamma
(-n_{3})\Gamma (-n_{4})}{\Gamma (\frac{D}{2}+n_{1}-n_{2})},  \label{f16}
\end{equation}%
where the dependent indexes (solutions of the system in $\left( \ref{l}%
\right) $ ) take the following values:

\begin{equation}
\begin{array}{l}
n_{2}=\frac{D}{2}-\nu _{1}-\nu _{2}-n_{1}, \\
\\
n_{3}=-\nu _{1}-n_{1}, \\
\\
n_{4}=-\nu _{2}-n_{1}.%
\end{array}%
\end{equation}%
Replacing now in $\left( \ref{f16}\right) $, we get:

\begin{equation}
G_{1}=(-1)^{-\frac{D}{2}}\dfrac{\left( -m^{2}\right) ^{\frac{D}{2}-\nu
_{1}+\nu _{2}}}{\tprod\nolimits_{j=1}^{2}\Gamma (\nu _{j})}%
\dsum\limits_{n_{1}}\dfrac{\left( p_{1}^{2}/m^{2}\right) ^{n_{1}}}{n_{1}!}%
\frac{\Gamma (\nu _{1}+\nu _{2}-\frac{D}{2}+n_{1})\Gamma (\nu
_{1}+n_{1})\Gamma (\nu _{2}+n_{1})}{\Gamma (\nu _{1}+\nu _{2}+2n_{1})},
\end{equation}%
and we can then reduce the above expression in terms of a hypergeometric
function:

\begin{equation}
G_{1}=\chi _{1}\;\left( -m^{2}\right) ^{\frac{D}{2}-\nu _{1}+\nu _{2}}\
_{3}F_{2}\left( \left.
\begin{array}{c}
\begin{array}{ccc}
\nu _{1}+\nu _{2}-\tfrac{D}{2}, & \nu _{1}, & \nu _{2}%
\end{array}
\\
\begin{array}{cc}
\tfrac{1}{2}+\tfrac{\nu _{1}+\nu _{2}}{2}, & \tfrac{\nu _{1}+\nu _{2}}{2}%
\end{array}%
\end{array}%
\right\vert \dfrac{p_{1}^{2}}{4m^{2}}\right) ,
\end{equation}%
where:

\begin{equation}
\chi _{1}=(-1)^{-\frac{D}{2}}\dfrac{\Gamma (\nu _{1}+\nu _{2}-\frac{D}{2})}{%
\Gamma (\nu _{1}+\nu _{2})}.
\end{equation}

\subsubsection{Serie representation when $n_{2}$\textit{\ is independent
index:}}

\qquad Analogously to the previous case, we now make the index $n_{2}$
independent, and from an adequate writing of the factors $\left\{ \Delta
_{i}\right\} $, we get the following expression associated to a free $n_{2}$:

\begin{equation}
G_{2}=\dfrac{(-1)^{-\frac{D}{2}}}{\tprod\nolimits_{j=1}^{2}\Gamma (\nu _{j})}%
\dsum\limits_{n_{2}}\dfrac{(-1)^{n_{2}}}{n_{2}!}\left( p_{1}^{2}\right)
^{n_{1}}\left( -m^{2}\right) ^{n_{2}}\frac{\Gamma (-n_{1})\Gamma
(-n_{3})\Gamma (-n_{4})}{\Gamma (\frac{D}{2}+n_{1}-n_{2})}.
\end{equation}%
This time the solutions for the dependent indexes are given by:

\begin{equation}
\begin{array}{l}
n_{1}=\frac{D}{2}-\nu _{1}-\nu _{2}-n_{2}, \\
\\
n_{3}=\nu _{2}-\frac{D}{2}+n_{2}, \\
\\
n_{4}=\nu _{1}-\frac{D}{2}+n_{2},%
\end{array}%
\end{equation}%
which allows to obtain the serie representation when $n_{2}$ is free in the
presolution $\left( \ref{f27}\right) $:

\begin{equation}
G_{2}=\chi _{2}\;\left( p_{1}^{2}\right) ^{\frac{D}{2}-\nu _{1}+\nu
_{2}}\;_{3}F_{2}\left( \left.
\begin{array}{c}
\begin{array}{ccc}
\nu _{1}+\nu _{2}-\tfrac{D}{2}, & \tfrac{1}{2}+\tfrac{\nu _{1}+\nu _{2}}{2}-%
\tfrac{D}{2}, & 1+\tfrac{\nu _{1}+\nu _{2}}{2}-\tfrac{D}{2}%
\end{array}
\\
\begin{array}{cc}
1+\nu _{1}-\tfrac{D}{2}, & 1+\nu _{2}-\tfrac{D}{2}%
\end{array}%
\end{array}%
\right\vert \dfrac{4m^{2}}{p_{1}^{2}}\right) ,
\end{equation}%
where the prefactor $\chi _{2}$ is given by the identity:

\begin{equation*}
\chi _{2}=(-1)^{-\frac{D}{2}}\dfrac{\Gamma (\nu _{1}+\nu _{2}-\frac{D}{2}%
)\Gamma (\frac{D}{2}-\nu _{1})\Gamma (\frac{D}{2}-\nu _{2})}{\Gamma (\nu
_{1})\Gamma (\nu _{2})\Gamma (D-\nu _{1}-\nu _{2})}.
\end{equation*}

\subsubsection{\textit{Solution for }$n_{3}$ and $n_{4}$\textit{\
independent:}}

\qquad Similar procedures to the previous ones allow us to quickly get the
terms $G_{3}$ and $G_{4}$ respectively:

\begin{equation}
G_{3}=\chi _{3}\;\left( p_{1}^{2}\right) ^{-\nu _{1}}\left( -m^{2}\right) ^{%
\frac{D}{2}-\nu _{2}}\;_{3}F_{2}\left( \left.
\begin{array}{c}
\begin{array}{ccc}
\nu _{1}, & \tfrac{1}{2}+\tfrac{\nu _{1}-\nu _{2}}{2}, & 1+\tfrac{\nu
_{1}-\nu _{2}}{2}%
\end{array}
\\
\begin{array}{cc}
1+\nu _{1}-\nu _{2}, & 1+\tfrac{D}{2}-\nu _{2}%
\end{array}%
\end{array}%
\right\vert \dfrac{4m^{2}}{p_{1}^{2}}\right) ,
\end{equation}%
and:%
\begin{equation}
G_{4}=\chi _{4}\;\left( p_{1}^{2}\right) ^{-\nu _{2}}\left( -m^{2}\right) ^{%
\frac{D}{2}-\nu _{1}}\;_{3}F_{2}\left( \left.
\begin{array}{c}
\begin{array}{ccc}
\nu _{2}, & \tfrac{1}{2}+\tfrac{\nu _{2}-\nu _{1}}{2}, & 1+\tfrac{\nu
_{2}-\nu _{1}}{2}%
\end{array}
\\
\begin{array}{cc}
1+\nu _{2}-\nu _{1}, & 1+\tfrac{D}{2}-\nu _{1}%
\end{array}%
\end{array}%
\right\vert \dfrac{4m^{2}}{p_{1}^{2}}\right) ,
\end{equation}%
where the following factors have been defined:

\begin{equation}
\chi _{3}=(-1)^{-\frac{D}{2}}\dfrac{\Gamma (\nu _{2}-\frac{D}{2})}{\Gamma
(\nu _{2})},
\end{equation}

\begin{equation}
\chi _{4}=(-1)^{-\frac{D}{2}}\dfrac{\Gamma (\nu _{1}-\frac{D}{2})}{\Gamma
(\nu _{1})}.
\end{equation}

\subsubsection{Solutions in the different kinematical regions, expressed as
sums of terms $G_{j}$}

\qquad We can distribute the previously found solutions in the two regions
where it is possible to expand the kinematical variables, the first located
in $\left\vert \dfrac{4m^{2}}{p_{1}^{2}}\right\vert <1$, where the solution
of $G$ is given by the expression:

\begin{equation}
G\left( \dfrac{4m^{2}}{p_{1}^{2}}\right) =G_{2}+G_{3}+G_{4},
\end{equation}%
and the solution in the region where $\left\vert \dfrac{p_{1}^{2}}{4m^{2}}%
\right\vert <1$:

\begin{equation}
\begin{array}{c}
\\
G\left( \dfrac{p_{1}^{2}}{4m^{2}}\right) =G_{1}.%
\end{array}%
\end{equation}%
In this way we have evaluated $G$ in terms of hypergeometric functions,
which correspond naturally to serie representations with respect to the two
energy scales present in diagram $G$.

\subsection{Example II : CBox diagram (On-Shell case)}

\qquad The next diagram has known solutions \cite{COl}, found by a different
method. An attempt using NDIM can be found in Ref. \cite{ASu15}.
Nevertheless, the result presented here is an improvement in the sense that
it is much simpler and easy to obtain. We do not need to use any
simplification tools in order to get the final result.

\begin{figure}[th]
\begin{center}
\epsfig{file=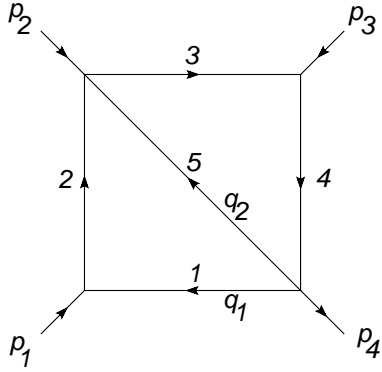,width=0.31\textwidth}
\end{center}
\caption{Labelled CBox diagram.}
\end{figure}
For this diagram (Fig. 3) the integral representation in momentum space is
given by:

\begin{equation}
G=\dint \frac{d^{D}q_{1}}{i\pi ^{\frac{D}{2}}}\frac{d^{D}q_{2}}{i\pi ^{\frac{%
D}{2}}}\frac{1}{(B_{1})^{\nu _{1}}}\frac{1}{(B_{2})^{\nu _{2}}}\frac{1}{%
(B_{3})^{\nu _{3}}}\frac{1}{(B_{4})^{\nu _{4}}}\frac{1}{(B_{5})^{\nu _{5}}},
\end{equation}%
where the quantities $B_{i}$ correspond to:

\begin{equation}
\begin{array}{l}
B_{1}=q_{1}^{2}-m_{1}^{2}+i0, \\
B_{2}=\left( q_{1}+p_{1}\right) ^{2}-m_{2}^{2}+i0, \\
B_{3}=\left( q_{1}+q_{2}+p_{1}+p_{2}\right) ^{2}-m_{3}^{2}+i0, \\
B_{4}=\left( q_{1}+q_{2}+p_{1}+p_{2}+p_{3}\right) ^{2}-m_{4}^{2}+i0, \\
B_{5}=q_{2}^{2}-m_{5}^{2}+i0.%
\end{array}%
\end{equation}%
According to the method developed in Ref. \cite{IGo}, we can find the
initial parameters matrix associated to the topology, which is:

\begin{equation}
\mathbf{M}=\left(
\begin{array}{ccccc}
x_{1}+x_{2}+x_{3}+x_{4} & x_{3}+x_{4} & x_{2}+x_{3}+x_{4} & x_{3}+x_{4} &
x_{4} \\
x_{3}+x_{4} & x_{3}+x_{4}+x_{5} & x_{3}+x_{4} & x_{3}+x_{4} & x_{4} \\
x_{2}+x_{3}+x_{4} & x_{3}+x_{4} & x_{2}+x_{3}+x_{4} & x_{3}+x_{4} & x_{4} \\
x_{3}+x_{4} & x_{3}+x_{4} & x_{3}+x_{4} & x_{3}+x_{4} & x_{4} \\
x_{4} & x_{4} & x_{4} & x_{4} & x_{4}%
\end{array}%
\right) .
\end{equation}%
Starting from this matrix we can find the algebraic components of the
parametric representation, evaluating for this purpose the polynomials $U$
and $F$. The multinomial $\left( 2-lineal\right) $ $U$ will be evaluated
using the determinant:

\begin{equation}
\begin{array}{ll}
U & =\left\vert
\begin{array}{cc}
x_{1}+x_{2}+x_{3}+x_{4} & x_{3}+x_{4} \\
x_{3}+x_{4} & x_{3}+x_{4}+x_{5}%
\end{array}%
\right\vert \\
&  \\
&
=x_{5}x_{1}+x_{5}x_{2}+x_{5}x_{3}+x_{5}x_{4}+x_{1}x_{3}+x_{1}x_{4}+x_{2}x_{3}+x_{2}x_{4},%
\end{array}%
\end{equation}%
and the multinomial $\left( 3-lineal\right) $ $F$ will be written as:

\begin{equation}
F=2C_{1,2}\;p_{1}.p_{2}+2C_{1,3}\;p_{1}.p_{3}+2C_{2,3}\;p_{2}.p_{3},
\end{equation}%
where the coefficients $C_{i,j}$ are evaluated in terms of subdeterminants
of the matrix of parameters, and then:

\begin{equation}
\begin{array}{l}
C_{1,2}=\left\vert
\begin{array}{ccc}
x_{1}+x_{2}+x_{3}+x_{4} & x_{3}+x_{4} & x_{2}+x_{3}+x_{4} \\
x_{3}+x_{4} & x_{3}+x_{4}+x_{5} & x_{3}+x_{4} \\
x_{3}+x_{4} & x_{3}+x_{4} & x_{3}+x_{4}%
\end{array}%
\right\vert =x_{1}x_{5}x_{3}+x_{1}x_{5}x_{4}, \\
\\
C_{1,3}=\left\vert
\begin{array}{ccc}
x_{1}+x_{2}+x_{3}+x_{4} & x_{3}+x_{4} & x_{2}+x_{3}+x_{4} \\
x_{3}+x_{4} & x_{3}+x_{4}+x_{5} & x_{3}+x_{4} \\
x_{4} & x_{4} & x_{4}%
\end{array}%
\right\vert =x_{1}x_{4}x_{5}, \\
\\
C_{2,3}=\left\vert
\begin{array}{ccc}
x_{1}+x_{2}+x_{3}+x_{4} & x_{3}+x_{4} & x_{3}+x_{4} \\
x_{3}+x_{4} & x_{3}+x_{4}+x_{5} & x_{3}+x_{4} \\
x_{4} & x_{4} & x_{4}%
\end{array}%
\right\vert =x_{4}x_{5}x_{1}+x_{4}x_{5}x_{2}.%
\end{array}%
\end{equation}%
We then have for $F$:

\begin{equation}
F=2\left( p_{1}.p_{2}\right) x_{1}x_{3}x_{5}\;+2\left( p_{2}.p_{3}\right)
x_{2}x_{4}x_{5}\;+2(p_{1}.p_{2}+p_{1}.p_{3}+p_{2}.p_{3})x_{1}x_{4}x_{5},
\end{equation}%
but since $%
2(p_{1}.p_{2}+p_{1}.p_{3}+p_{2}.p_{3})=(p_{1}+p_{2}+p_{3})^{2}=p_{4}^{2}=0$,
$2p_{1}.p_{2}=(p_{1}+p_{2})^{2}=s$ and $2p_{2}.p_{3}=(p_{2}+p_{3})^{2}=t$,
we can rewrite $F$ as:

\begin{equation}
F=x_{1}x_{3}x_{5}\;s+x_{2}x_{4}x_{5}\;t.
\end{equation}

\subsubsection{Massless case $\left( m_{1}=...=m_{5}=0\right) $}

\qquad The simplest case is the one that does not consider mass scales in
the diagram. In this situation the parametric representation of $G$ has the
form:

\begin{equation}
G=\dfrac{(-1)^{-D}}{\tprod\nolimits_{j=1}^{5}\Gamma (\nu _{j})}%
\dint\limits_{0}^{\infty }d\overrightarrow{x}\;\frac{\exp \left( -\dfrac{%
x_{1}x_{3}x_{5}}{U}s\right) \exp \left( -\dfrac{x_{2}x_{4}x_{5}}{U}t\right)
}{U^{\frac{D}{2}}}.
\end{equation}%
The first step is the expansion of the exponentials, after which we find:

\begin{equation}
G=\dfrac{(-1)^{-D}}{\tprod\nolimits_{j=1}^{5}\Gamma (\nu _{j})}%
\dsum\limits_{n_{1},n_{2}}\phi _{n_{1},n_{2}}\;\left( s\right)
^{n_{1}}\left( t\right) ^{n_{2}}\dint\limits_{0}^{\infty }d\overrightarrow{x}%
\;\dfrac{%
x_{1}^{n_{1}}x_{3}^{n_{1}}x_{2}^{n_{2}}x_{4}^{n_{2}}x_{5}^{n_{1}+n_{2}}}{U^{%
\frac{D}{2}+n_{1}+n_{2}}}.  \label{f17}
\end{equation}%
The exponentials are not susceptible to factorization, since the exponents
contain only one term. On the other hand, in order to apply the method the
polynomial $U$ is factorized in the following form:

\begin{equation}
U=f_{1}f_{2}+f_{1}x_{5}+f_{2}x_{5},
\end{equation}%
where we have defined the submultinomials:

\begin{equation}
\begin{array}{l}
f_{1}=(x_{1}+x_{2}), \\
f_{2}=(x_{3}+x_{4}).%
\end{array}%
\end{equation}%
We now proceed to find the multiregion expansion for the factorized
multinomial $U$, which reads:

\begin{equation}
\begin{array}{ll}
\dfrac{1}{U^{\frac{D}{2}+n_{1}+n_{2}}} & =\dfrac{1}{\left[
f_{1}f_{2}+f_{1}x_{5}+f_{2}x_{5}\right] ^{\frac{D}{2}+n_{1}+n_{2}}} \\
&  \\
& =\dsum\limits_{n_{3},..,n_{5}}\phi _{n_{3},..,n_{5}}\;\dfrac{\left(
f_{1}\right) ^{n_{3}+n_{4}}\left( f_{2}\right)
^{n_{3}+n_{5}}x_{5}^{n_{4}+n_{5}}}{\Gamma (\frac{D}{2}+n_{1}+n_{2})}\Delta
_{1},%
\end{array}%
\end{equation}%
and the expansion of the submultinomials $f_{1}$ and $f_{2}$:

\begin{equation}
\begin{array}{l}
\left( f_{1}\right)
^{n_{3}+n_{4}}=(x_{1}+x_{2})^{n_{3}+n_{4}}=\dsum\limits_{n_{6},n_{7}}\phi
_{n_{6},n_{7}}\;x_{1}^{n_{6}}x_{2}^{n_{7}}\dfrac{\Delta _{2}}{\Gamma
(-n_{3}-n_{4})}, \\
\\
\left( f_{2}\right)
^{n_{3}+n_{5}}=(x_{3}+x_{4})^{n_{3}+n_{5}}=\dsum\limits_{n_{8},n_{9}}\phi
_{n_{8},n_{9}}\;x_{3}^{n_{8}}x_{4}^{n_{9}}\dfrac{\Delta _{3}}{\Gamma
(-n_{3}-n_{5})},%
\end{array}%
\end{equation}%
where the costraints $\left\{ \Delta _{i}\right\} $ are given by the
following equations:

\begin{equation}
\begin{array}{l}
\Delta _{1}=\left\langle \frac{D}{2}+n_{1}+n_{2}+n_{3}+n_{4}+n_{5}\right%
\rangle , \\
\Delta _{2}=\left\langle -n_{3}-n_{4}+n_{6}+n_{7}\right\rangle , \\
\Delta _{3}=\left\langle -n_{3}-n_{5}+n_{8}+n_{9}\right\rangle .%
\end{array}%
\end{equation}%
Finally the multiregion expansion for $U$ can be written as follows:

\begin{equation}
\dfrac{1}{U^{\frac{D}{2}+n_{1}+n_{2}}}=\dsum\limits_{n_{3},..,n_{9}}\phi
_{n_{3},..,n_{9}}\;\dfrac{%
x_{1}^{n_{6}}x_{2}^{n_{7}}x_{3}^{n_{8}}x_{4}^{n_{9}}x_{5}^{n_{4}+n_{5}}}{%
\Gamma (\frac{D}{2}+n_{1}+n_{2})}\frac{\tprod\nolimits_{j=1}^{3}\Delta _{j}}{%
\Gamma (-n_{3}-n_{4})\Gamma (-n_{3}-n_{5})},
\end{equation}%
then replacing in $\left( \ref{f17}\right) $ and separating the integration
variables, we get:

\begin{equation}
\begin{array}{ll}
G= & \dfrac{(-1)^{-D}}{\tprod\nolimits_{j=1}^{5}\Gamma (\nu _{j})}%
\dsum\limits_{n_{1},..,n_{9}}\phi _{n_{1},..,n_{9}}\;\dfrac{\left( s\right)
^{n_{1}}\left( t\right) ^{n_{2}}}{\Gamma (\frac{D}{2}+n_{1}+n_{2})}\dfrac{%
\tprod\nolimits_{j=1}^{3}\Delta _{j}}{\Gamma (-n_{3}-n_{4})\Gamma
(-n_{3}-n_{5})} \\
&  \\
& \dint dx_{1\;}x_{1}^{n_{1}+n_{6}}\dint dx_{2\;}x_{2}^{n_{2}+n_{7}}\dint
dx_{3\;}x_{3}^{n_{1}+n_{8}}\dint dx_{4\;}x_{4}^{n_{2}+n_{9}}\dint
dx_{5\;}x_{5}^{n_{1}+n_{2}+n_{4}+n_{5}},%
\end{array}%
\end{equation}%
and changing the integrations by their equivalents $\left\langle \cdot
\right\rangle $, we finally obtain the presolution of $G$:

\begin{equation}
G=\dfrac{(-1)^{-D}}{\tprod\nolimits_{j=1}^{5}\Gamma (\nu _{j})}%
\dsum\limits_{n_{1},..,n_{9}}\phi _{n_{1},..,n_{9}}\;\dfrac{\left( s\right)
^{n_{1}}\left( t\right) ^{n_{2}}}{\Gamma (\frac{D}{2}+n_{1}+n_{2})}\dfrac{%
\tprod\nolimits_{j=1}^{8}\Delta _{j}}{\Gamma (-n_{3}-n_{4})\Gamma
(-n_{3}-n_{5})},
\end{equation}%
where the constrains $\left\{ \Delta _{i}\right\} $\ associated to the
integrals are:

\begin{equation}
\begin{array}{l}
\Delta _{4}=\left\langle \nu _{1}+n_{1}+n_{6}\right\rangle , \\
\Delta _{5}=\left\langle \nu _{2}+n_{2}+n_{7}\right\rangle , \\
\Delta _{6}=\left\langle \nu _{3}+n_{1}+n_{8}\right\rangle , \\
\Delta _{7}=\left\langle \nu _{4}+n_{2}+n_{9}\right\rangle , \\
\Delta _{8}=\left\langle \nu _{5}+n_{1}+n_{2}+n_{4}+n_{5}\right\rangle .%
\end{array}%
\end{equation}%
In this case we generate at most $C_{8}^{9}=9$ terms or contributions to the
solution of $G$, each of which has the form $\ _{l}F_{\left( l-1\right) }$.
Nevertheless, three of them do not contribute to the solution, and the
remaining terms are distributed in two kinematical regions, which we now
summarize.

\paragraph{Solutions in the region $\left\vert \dfrac{s}{t}\right\vert <1 $:}

\begin{equation}
G\left( \dfrac{s}{t}\right) =G_{1}+G_{7}+G_{9}.
\end{equation}

\paragraph{Solutions in the region $\left\vert \dfrac{t}{s}\right\vert <1 $:}

\begin{equation}
G\left( \dfrac{t}{s}\right) =G_{2}+G_{6}+G_{8}.
\end{equation}

\paragraph{Explicit analytical solution for the kinematical region $%
\left\vert \dfrac{s}{t}\right\vert <1$.}

\qquad In order to compare our results with those of Ref. \cite{COl}, we now
present the solution which corresponds to the limit $\left\vert \dfrac{s}{t}%
\right\vert <1$. The contributions that correspond to the solution of the
diagram in this region are associated to the free indexes $n_{1}$, $n_{7}$
and $n_{9}$. In order to simplify the final resulting expressions we have
used the following notation $\nu _{ijk...}=\nu _{i}+\nu _{j}+\nu _{k}+...$.

Explicitly the solution in this kinematical region can be written as:

\begin{equation}
G\left( \dfrac{s}{t}\right) =G_{1}+G_{7}+G_{9},
\end{equation}%
where the contributions $G_{1}$, $G_{7}$, $G_{9}$ are given by:
\begin{equation}
G_{1}=\chi _{1}\;t^{D-\nu _{12345}}\;_{3}F_{2}\left( \left.
\begin{array}{c}
\nu _{1},\quad \nu _{3,}\quad \nu _{12345}-D \\
1+\nu _{1345}-D,\quad 1+\nu _{1235}-D%
\end{array}%
\right\vert -\dfrac{s}{t}\right) ,
\end{equation}%
and the prefactor $\chi _{1}$\ corresponds to:

\begin{equation}
\chi _{1}=(-1)^{-D}\dfrac{\Gamma (\nu _{12345}-D)\Gamma (\frac{D}{2}-\nu
_{12})\Gamma (\frac{D}{2}-\nu _{34})\Gamma (\frac{D}{2}-\nu _{5})\Gamma
(D-\nu _{1345})\Gamma (D-\nu _{1235})}{\Gamma (\nu _{2})\Gamma (\nu
_{4})\Gamma (\nu _{5})\Gamma (\frac{3D}{2}-\nu _{12345})\Gamma (D-\nu
_{125})\Gamma (D-\nu _{345})}.
\end{equation}%
We also have:

\begin{equation}
G_{7}=\chi _{7}\;s^{D-\nu _{1345}}t^{-\nu _{2}}\;_{3}F_{2}\left( \left.
\begin{array}{c}
\nu _{2},\quad D-\nu _{345},\quad D-\nu _{145} \\
1+D-\nu _{1345},\quad 1-\nu _{4}+\nu _{2}%
\end{array}%
\right\vert -\dfrac{s}{t}\right) ,
\end{equation}%
where the prefactor $\chi _{7}$ is:

\begin{equation}
\chi _{7}=(-1)^{-D}\dfrac{\Gamma (\frac{D}{2}-\nu _{12})\Gamma (\frac{D}{2}%
-\nu _{34})\Gamma (\frac{D}{2}-\nu _{5})\Gamma (D-\nu _{145})\Gamma (\nu
_{1345}-D)\Gamma (\nu _{4}-\nu _{2})}{\Gamma (\nu _{1})\Gamma (\nu
_{3})\Gamma (\nu _{4})\Gamma (\nu _{5})\Gamma (\frac{3D}{2}-\nu
_{12345})\Gamma (D-\nu _{125})},
\end{equation}%
and finally:

\begin{equation}
G_{9}=\chi _{9}\;s^{D-\nu _{1235}}t^{-\nu _{4}}\;_{3}F_{2}\left( \left.
\begin{array}{c}
\nu _{4},\quad D-\nu _{235},\quad D-\nu _{125} \\
1+D-\nu _{1235},\quad 1-\nu _{2}+\nu _{4}%
\end{array}%
\right\vert -\dfrac{s}{t}\right) ,
\end{equation}%
with:

\begin{equation}
\chi _{9}=(-1)^{-D}\dfrac{\Gamma (\nu _{1235}-D)\Gamma (\frac{D}{2}-\nu
_{12})\Gamma (\frac{D}{2}-\nu _{34})\Gamma (\frac{D}{2}-\nu _{5})\Gamma
(D-\nu _{235})\Gamma (\nu _{2}-\nu _{4})}{\Gamma (\nu _{1})\Gamma (\nu
_{2})\Gamma (\nu _{3})\Gamma (\nu _{5})\Gamma (\frac{3D}{2}-\nu
_{12345})\Gamma (D-\nu _{345})}.
\end{equation}

\subsubsection{Massive CBox diagram}

\qquad The next level of difficulty of this problem corresponds to the
addition of another energy scale in the topology. In particular let us
consider associating masses of magnitude $m$ to some of the propagators of
the diagram. As a result we now obtain two-variable hypergeometric functions
as solutions.

There are different alternatives for factorizing the multinomial $U$ and
extending the solution of the massless case found previously to a set of
massive cases. Nevertheless, not all the possible distributions of the mass
scale in the propagators generate two-variable series, and in general an
arbitrary assignment of masses, even if they are equal, produces series
solutions of multiplicity $\mu >2$. The following forms of factorization of $%
U$ allow us to visualize any cases in which it is possible to solve the
problem in terms of two-variable series.

\paragraph{Factorization I.}

\begin{equation}
U=f_{1}f_{2}+f_{1}x_{5}+f_{2}x_{5},  \label{f41}
\end{equation}%
where we have defined the submultinomials:

\begin{equation}
\begin{array}{l}
f_{1}=(x_{1}+x_{2}), \\
f_{2}=(x_{3}+x_{4}).%
\end{array}%
\end{equation}%
With this factorization it is possible to have a two-variable solution for
the following massive cases:

a)

\begin{equation}
\begin{array}{l}
m_{1}=m_{2}=m, \\
m_{3}=m_{4}=m_{5}=0.%
\end{array}%
\end{equation}

\begin{equation*}
\text{or}
\end{equation*}

b)

\begin{equation}
\begin{array}{l}
m_{1}=m_{2}=m_{5}=0, \\
m_{3}=m_{4}=m.%
\end{array}%
\end{equation}

\paragraph{Factorization II.}

Another possible factorization is given by the expression:

\begin{equation}
U=f_{2}x_{1}+f_{2}x_{2}+f_{1}x_{5},  \label{f40}
\end{equation}%
where the submultinomials $f_{i}$ are given by:

\begin{equation}
\begin{array}{l}
f_{2}=(f_{1}+x_{5}), \\
f_{1}=(x_{3}+x_{4}),%
\end{array}%
\end{equation}%
which allows the evaluation of a more complex case, containing the following
mass distribution:

\begin{equation}
\begin{array}{l}
m_{1}=m_{2}=0, \\
m_{3}=m_{4}=m_{5}=m.%
\end{array}%
\end{equation}

\paragraph{Factorization III.}

We can have yet another factorization of $U$:

\begin{equation}
U=f_{2}x_{3}+f_{2}x_{4}+f_{1}x_{5},  \label{f18}
\end{equation}%
where we have defined:

\begin{equation}
\begin{array}{l}
f_{2}=(f_{1}+x_{5}), \\
f_{1}=(x_{1}+x_{2}).%
\end{array}%
\end{equation}%
Such a factorization allows to consider the case when we have the following
configuration:

\begin{equation}
\begin{array}{l}
m_{1}=m_{2}=m_{5}=m, \\
m_{3}=m_{4}=0.%
\end{array}%
\end{equation}

\paragraph{Obtaining the presolution in a case with massive propagators.}

\qquad In order to show once again the simplicity of NDIM, we will find the
multiregion expansion of the diagram $G$ associated to the following
particular distribution of masses $\left( m_{1}=m_{2}=m_{5}=m\right) $ and $%
\left( m_{3}=m_{4}=0\right) $, and we will show explicitly the solution
associated to a specific kinematical region. In this case the integral
representation in terms of Schwinger parameters is given by:

\begin{equation}
G=\dfrac{(-1)^{-D}}{\tprod\nolimits_{j=1}^{5}\Gamma (\nu _{j})}%
\dint\limits_{0}^{\infty }d\overrightarrow{x}\;\frac{\exp \left(
-f_{2}\left( -m^{2}\right) \right) \exp \left( -\dfrac{x_{1}x_{3}x_{5}}{U}%
s\right) \exp \left( -\dfrac{x_{2}x_{4}x_{5}}{U}t\right) }{U^{\frac{D}{2}}},
\end{equation}%
where the multinomial $U$ is given by $\left( \ref{f18}\right) $:

\begin{equation}
U=f_{2}x_{3}+f_{2}x_{4}+f_{1}x_{5},
\end{equation}%
with the submultinomials:

\begin{equation}
\begin{array}{ccc}
f_{2}=(f_{1}+x_{5}) & \text{and} & f_{1}=(x_{1}+x_{2}).%
\end{array}%
\end{equation}%
The expansion of the exponentials gives us:

\begin{equation}
G=\dfrac{(-1)^{-D}}{\tprod\nolimits_{j=1}^{5}\Gamma (\nu _{j})}%
\dsum\limits_{n_{1},..,n_{3}}\phi _{n_{1},..,n_{3}}\;\left( -m^{2}\right)
^{n_{1}}\left( s\right) ^{n_{2}}\left( t\right)
^{n_{3}}\dint\limits_{0}^{\infty }d\overrightarrow{x}\;\dfrac{%
x_{1}^{n_{2}}x_{2}^{n_{3}}x_{3}^{n_{2}}x_{4}^{n_{3}}x_{5}^{n_{2}+n_{3}}}{U^{%
\frac{D}{2}+n_{2}+n_{3}}}\left( f_{2}\right) ^{n_{1}},
\end{equation}%
where the order in which the multiregion expansions to the multinomials that
are present ($U,\;f_{2}$ and $f_{1}$) have to be made is simple, and is the
following, according to the dependence level between them:

\begin{equation*}
U\left( f_{2},f_{1}\right) \longrightarrow f_{2}\left( f_{1}\right)
\longrightarrow f_{1}.
\end{equation*}%
A little algebra allows to finally obtain the presolution of the diagram $G$%
. Explicitly we have that:

\begin{equation}
G=\dfrac{(-1)^{-D}}{\tprod\nolimits_{j=1}^{5}\Gamma (\nu _{j})}%
\dsum\limits_{n_{1},..,n_{10}}\phi _{n_{1},..,n_{10}}\;\dfrac{\left(
-m^{2}\right) ^{n_{1}}\left( s\right) ^{n_{2}}\left( t\right) ^{n_{3}}}{%
\Gamma (\frac{D}{2}+n_{2}+n_{3})}\dfrac{\tprod\nolimits_{j=1}^{8}\Delta _{j}%
}{\Gamma (-n_{1}-n_{4}-n_{5})\Gamma (-n_{6}-n_{7})},
\end{equation}%
where the following constraints have been defined:

\begin{equation}
\begin{array}{l}
\Delta _{1}=\left\langle \frac{D}{2}+n_{2}+n_{3}+n_{4}+n_{5}+n_{6}\right%
\rangle , \\
\Delta _{2}=\left\langle -n_{1}-n_{4}-n_{5}+n_{7}+n_{8}\right\rangle , \\
\Delta _{3}=\left\langle -n_{6}-n_{7}+n_{9}+n_{10}\right\rangle , \\
\Delta _{4}=\left\langle \nu _{1}+n_{2}+n_{9}\right\rangle , \\
\Delta _{5}=\left\langle \nu _{2}+n_{3}+n_{10}\right\rangle , \\
\Delta _{6}=\left\langle \nu _{3}+n_{2}+n_{4}\right\rangle , \\
\Delta _{7}=\left\langle \nu _{4}+n_{3}+n_{5}\right\rangle , \\
\Delta _{8}=\left\langle \nu _{5}+n_{2}+n_{3}+n_{6}+n_{8}\right\rangle .%
\end{array}%
\end{equation}%
According to this result we can say that the total number of possible
contributions to the solution is $C_{8}^{10}=45$, and that evidently they
correspond to series of multiplicity two. Nevertheless, 18 of these do not
really contribute due to the particular nature of the linear system build
from the constraints. We now show the remaining contributions, in terms of a
set of components $G_{i,j}$, according to the type of two-variable
hypergeometric function ($F^{\substack{ p:r:u  \\ q:s:v}}$ or $\overline{F}
^{\substack{ p:r:u  \\ q:s:v}}$ , see appendix) that is generated and also
according to the arguments of this same function:

\bigskip

\begin{itemize}
\item[Set. 1 :] $G_{1,5}+G_{1,10}+G_{7,10}+G_{8,10}\Longrightarrow \overline{%
F}^{\substack{ 1:4:1  \\ 2:2:1}}\left( \left.
\begin{array}{cc}
&
\end{array}%
\right\vert \dfrac{4m^{2}}{s},\dfrac{s}{t}\right) $

\item[Set. 2 :] $G_{5,7}+G_{5,8}\Longrightarrow \overline{F}^{\substack{ %
2:1:4  \\ 1:1:2}}\left( \left.
\begin{array}{cc}
&
\end{array}%
\right\vert -\dfrac{s}{t},-\dfrac{4m^{2}}{s}\right) $

\item[Set. 3 :] $G_{1,4}+G_{1,9}+G_{7,9}+G_{8,9}\Longrightarrow \overline{F}
^{\substack{ 1:4:1  \\ 2:2:1}}\left( \left.
\begin{array}{cc}
&
\end{array}%
\right\vert \dfrac{4m^{2}}{t},\dfrac{t}{s}\right) $

\item[Set. 4 :] $G_{4,7}+G_{4,8}\Longrightarrow \overline{F}^{\substack{ %
2:1:4  \\ 1:1:2}}\left( \left.
\begin{array}{cc}
&
\end{array}%
\right\vert -\dfrac{t}{s},-\dfrac{4m^{2}}{t}\right) $

\item[Set. 5 :] $G_{4,5}+G_{4,10}+G_{5,9}+\left( G_{9,10}=0\right)
\Longrightarrow F^{\substack{ 4:1:1  \\ 3:1:1}}\left( \left.
\begin{array}{cc}
&
\end{array}%
\right\vert \dfrac{4m^{2}}{s},\dfrac{4m^{2}}{t}\right) $

\item[Set. 6 :] $G_{1,3}+G_{3,7}+G_{3,8}\Longrightarrow F^{\substack{ 1:4:2
\\ 2:2:0}}\left( \left.
\begin{array}{cc}
&
\end{array}%
\right\vert \dfrac{4m^{2}}{s},-\dfrac{t}{s}\right) $

\item[Set. 7 :] $G_{1,2}+G_{2,7}+G_{2,8}\Longrightarrow F^{\substack{ 1:4:2
\\ 2:2:0}}\left( \left.
\begin{array}{cc}
&
\end{array}%
\right\vert \dfrac{4m^{2}}{t},-\dfrac{s}{t}\right) $

\item[Set. 8 :] $G_{2,5}+G_{2,10}\Longrightarrow \overline{F}^{\substack{ %
3:2:1  \\ 4:0:1}}\left( \left.
\begin{array}{cc}
&
\end{array}%
\right\vert \dfrac{s}{4m^{2}},-\dfrac{4m^{2}}{t}\right) $

\item[Set. 9 :] $G_{3,4}+G_{3,9}\Longrightarrow \overline{F}^{\substack{ %
3:2:1  \\ 4:0:1}}\left( \left.
\begin{array}{cc}
&
\end{array}%
\right\vert \dfrac{t}{4m^{2}},-\dfrac{4m^{2}}{s}\right) $

\item[Set. 10 :] $G_{2,3}\Longrightarrow F^{\substack{ 3:2:2  \\ 4:0:0}}%
\left( \left.
\begin{array}{cc}
&
\end{array}%
\right\vert \dfrac{s}{4m^{2}},\dfrac{t}{4m^{2}}\right) .$
\end{itemize}

\bigskip

It is important to notice that each set of contributions does not
necessarily constitutes the final solution in certain kinematical region,
because it is possible that for some cases the type of solutions get mixed.
As an example let us consider the case where the magnitude of the variable $%
\left\vert t\right\vert $ is the dominant, in which case and according to
the convergence conditions of each set of solutions, it will be possible to
write immediately the solution in this region as the algebraic sum of the
sets 1, 2, 5, 7 and 8.

Now let us write explicitly one of the solutions of $G$. For this purpose we
consider the solutions that are present in the region where $\left(
4m^{2}>\left\vert s\right\vert \right) $ and $\left( 4m^{2}>\left\vert
t\right\vert \right) $, which according to the previous set is composed of
only one contribution, $G_{2,3}$:

\begin{equation}
G\left( \dfrac{s}{4m^{2}},\dfrac{t}{4m^{2}}\right) =G_{2,3},
\end{equation}%
where the contribution $G_{2,3}$ refers to the series:

\begin{equation}
G_{2,3}=\dfrac{(-1)^{-D}}{\tprod\nolimits_{j=1}^{5}\Gamma (\nu _{j})}%
\dsum\limits_{n_{2},n_{3}}\frac{(-1)^{n_{2}+n_{3}}}{n_{2}!n_{3}!}\dfrac{%
\left( -m^{2}\right) ^{n_{1}}\left( s\right) ^{n_{2}}\left( t\right) ^{n_{3}}%
}{\Gamma (D/2+n_{2}+n_{3})}\dfrac{\tprod\nolimits_{\substack{ j=1,  \\ j\neq
2,3}}^{8}\Gamma (-n_{j})}{\Gamma (-n_{1}-n_{4}-n_{5})\Gamma (-n_{6}-n_{7})},
\label{f19}
\end{equation}%
and where the dependent indexes take the following values:

\begin{equation}
\begin{array}{l}
n_{1}=-\nu _{12345}+D-n_{2}-n_{3}, \\
n_{4}=-\nu _{3}-n_{2}, \\
n_{5}=-\nu _{4}-n_{3}, \\
n_{6}=\nu _{34}-\frac{D}{2}, \\
n_{7}=-\nu _{1234}+\frac{D}{2}-n_{2}-n_{3}, \\
n_{8}=-\nu _{345}+\frac{D}{2}-n_{2}-n_{3}, \\
n_{9}=-\nu _{1}-n_{2}, \\
n_{10}=\nu _{2}-n_{3}.%
\end{array}%
\end{equation}%
Making the corresponding replacements and after a little algebra in the
expression $\left( \ref{f19}\right) $, we obtain the representation in terms
of the Kamp\'{e} de F\'{e}riet generalized hypergeometric function $\left( F
^{\substack{ p:r:u  \\ q:s:v}}\right) $:

\begin{equation}
G\left( \dfrac{s}{4m^{2}},\dfrac{t}{4m^{2}}\right) =\chi \;(-m^{2})^{D-\nu
_{12345}}\;F^{\substack{ 3:2:2  \\ 4:0:0}}\left( \left.
\begin{array}{ccc}
\left\{ \alpha \right\} & \left\{ a\right\} & \left\{ c\right\} \\
\left\{ \beta \right\} & \left\{ -\right\} & \left\{ -\right\}%
\end{array}%
\right\vert
\begin{array}{c}
\dfrac{s}{4m^{2}},\dfrac{t}{4m^{2}}%
\end{array}%
\right) ,
\end{equation}%
where the prefactor $\chi $ is given by:

\begin{equation}
\chi =(-1)^{-D}\frac{\Gamma (\nu _{12345}-D)\Gamma (\nu _{1234}-\frac{D}{2}%
)\Gamma (\nu _{345}-\frac{D}{2})\Gamma (\frac{D}{2}-\nu _{34})}{\Gamma (\nu
_{5})\Gamma (\nu _{125}+2\nu _{34}-D)\Gamma (\nu _{12})\Gamma (\frac{D}{2})},
\end{equation}%
and where the corresponding parameters are:

\begin{equation}
\begin{array}{lll}
\alpha _{1}=\nu _{12345}-D, &  & \beta _{4}=\frac{1}{2}+\tfrac{\nu _{125}}{2}%
+\nu _{34}-\frac{D}{2}, \\
\alpha _{2}=\nu _{1234}-\frac{D}{2}, &  & a_{1}=\nu _{1}, \\
\alpha _{3}=\nu _{345}-\frac{D}{2}, &  & a_{2}=\nu _{3}, \\
\beta _{1}=\nu _{12}, &  & c_{1}=\nu _{2}, \\
\beta _{2}=\frac{D}{2}, &  & c_{2}=\nu _{4}, \\
\beta _{3}=\tfrac{\nu _{125}}{2}+\nu _{34}-\frac{D}{2}. &  &
\end{array}%
\end{equation}%
In this case the series converges if the following condition is satisfied:

\begin{equation}
\begin{array}{c}
\max \left\{ \left\vert \dfrac{s}{4m^{2}}\right\vert ,\left\vert \dfrac{t}{%
4m^{2}}\right\vert \right\} <1.%
\end{array}%
\end{equation}

\subsection{Example III : Four-loop propagator}

\qquad In this example the full power of the integration technique here
presented is revealed, due to the fact that the parametric integral that
represents the diagram of Fig. 4 is quite complex.

\begin{figure}[ht]
\begin{center}
\epsfig{file=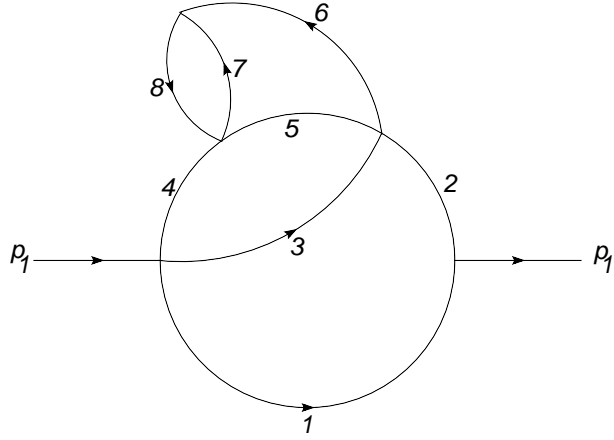,width=0.5\textwidth}
\end{center}
\caption{Labelled four-loop propagator.}
\end{figure}
The integral representation in momentum space is in this case:

\begin{equation}
G=\dint \frac{d^{D}q_{1}}{i\pi ^{\frac{D}{2}}}...\frac{d^{D}q_{4}}{i\pi ^{%
\frac{D}{2}}}\;\tprod\limits_{j=1}^{8}\frac{1}{(k_{j}^{2}-m_{j}^{2}+i0)^{\nu
_{j}}},
\end{equation}%
where the branch momenta $k_{i}$ $(i=1,...,8)$ are respectively:

\begin{equation}
\begin{array}{l}
k_{1}=q_{1}+p_{1}, \\
k_{2}=q_{1}, \\
k_{3}=q_{2}-q_{1}, \\
k_{4}=q_{2}, \\
k_{5}=q_{2}-q_{3}, \\
k_{6}=q_{3}, \\
k_{7}=q_{4}-q_{3}, \\
k_{8}=q_{4}.%
\end{array}%
\end{equation}%
The initial parameter matrix associated to the diagram can be easily
obtained from its topology:

\begin{equation}
\mathbf{M}=\left(
\begin{array}{ccccc}
x_{1}+x_{2}+x_{3} & -x_{3} & 0 & 0 & x_{1} \\
-x_{3} & x_{3}+x_{4}+x_{5} & -x_{5} & 0 & 0 \\
0 & -x_{5} & x_{5}+x_{6}+x_{7} & -x_{7} & 0 \\
0 & 0 & -x_{7} & x_{7}+x_{8} & 0 \\
x_{1} & 0 & 0 & 0 & x_{1}%
\end{array}%
\right) .  \label{f6}
\end{equation}%
The multinomial $U$ is obtained from the evaluation of the following
determinant:

\begin{equation}
\begin{array}{ll}
U= & \left\vert
\begin{array}{cccc}
x_{1}+x_{2}+x_{3} & -x_{3} & 0 & 0 \\
-x_{3} & x_{3}+x_{4}+x_{5} & -x_{5} & 0 \\
0 & -x_{5} & x_{5}+x_{6}+x_{7} & -x_{7} \\
0 & 0 & -x_{7} & x_{7}+x_{8}%
\end{array}%
\right\vert \\
&  \\
U= &
x_{1}x_{3}x_{5}x_{7}+x_{1}x_{3}x_{5}x_{8}+x_{1}x_{3}x_{6}x_{7}+x_{1}x_{4}x_{5}x_{7}+x_{2}x_{3}x_{5}x_{7}+
\\
&
x_{1}x_{3}x_{6}x_{8}+x_{1}x_{4}x_{5}x_{8}+x_{1}x_{4}x_{6}x_{7}+x_{2}x_{3}x_{5}x_{8}+x_{2}x_{3}x_{6}x_{7}+
\\
&
x_{2}x_{4}x_{5}x_{7}+x_{1}x_{3}x_{7}x_{8}+x_{1}x_{4}x_{6}x_{8}+x_{1}x_{5}x_{6}x_{7}+x_{2}x_{3}x_{6}x_{8}+
\\
&
x_{2}x_{4}x_{5}x_{8}+x_{2}x_{4}x_{6}x_{7}+x_{3}x_{4}x_{5}x_{7}+x_{1}x_{4}x_{7}x_{8}+x_{1}x_{5}x_{6}x_{8}+
\\
&
x_{2}x_{3}x_{7}x_{8}+x_{2}x_{4}x_{6}x_{8}+x_{2}x_{5}x_{6}x_{7}+x_{3}x_{4}x_{5}x_{8}+x_{3}x_{4}x_{6}x_{7}+
\\
&
x_{1}x_{5}x_{7}x_{8}+x_{2}x_{4}x_{7}x_{8}+x_{2}x_{5}x_{6}x_{8}+x_{3}x_{4}x_{6}x_{8}+x_{3}x_{5}x_{6}x_{7}+
\\
&
x_{2}x_{5}x_{7}x_{8}+x_{3}x_{4}x_{7}x_{8}+x_{3}x_{5}x_{6}x_{8}+x_{3}x_{5}x_{7}x_{8},%
\end{array}%
\end{equation}%
and $F$ is in turn obtained from the evaluation of the determinant of the
expression $\left( \ref{f6}\right) $:

\begin{equation}
\begin{array}{ll}
F= &
(x_{1}x_{2}x_{3}x_{5}x_{7}+x_{1}x_{2}x_{3}x_{5}x_{8}+x_{1}x_{2}x_{3}x_{6}x_{7}+x_{1}x_{2}x_{4}x_{5}x_{7}+
\\
&
x_{1}x_{2}x_{3}x_{6}x_{8}+x_{1}x_{2}x_{4}x_{5}x_{8}+x_{1}x_{2}x_{4}x_{6}x_{7}+x_{1}x_{3}x_{4}x_{5}x_{7}+
\\
&
x_{1}x_{2}x_{3}x_{7}x_{8}+x_{1}x_{2}x_{4}x_{6}x_{8}+x_{1}x_{2}x_{5}x_{6}x_{7}+x_{1}x_{3}x_{4}x_{5}x_{8}+
\\
&
x_{1}x_{3}x_{4}x_{6}x_{7}+x_{1}x_{2}x_{4}x_{7}x_{8}+x_{1}x_{2}x_{5}x_{6}x_{8}+x_{1}x_{3}x_{4}x_{6}x_{8}+
\\
&
x_{1}x_{3}x_{5}x_{6}x_{7}+x_{1}x_{2}x_{5}x_{7}x_{8}+x_{1}x_{3}x_{4}x_{7}x_{8}+x_{1}x_{3}x_{5}x_{6}x_{8}+
\\
& x_{1}x_{3}x_{5}x_{7}x_{8})\;p_{1}^{2}.%
\end{array}%
\end{equation}

\subsubsection{Factorization of the multinomials $U$ and $F$}

\qquad The large number of terms that $U(34)$ and $F(21)$ have makes
impossible to get manageable solutions if the method is applied without
performing the necessary factorizations first. Nevertheless, this type of
diagrams that in the massless case are solvable loop by loop, present a form
such that they can be easily factorized in submultinomials. For this
particular case the adequate multimomial factorization is:

\begin{equation}
\begin{array}{l}
F=x_{1}f_{7}\;p_{1}^{2}, \\
\\
U=x_{1}f_{6}+f_{7},%
\end{array}%
\end{equation}%
where the functions $f_{i}$ are given by the equations:

\begin{equation}
\begin{array}{l}
f_{7}=(x_{2}f_{6}+f_{5}), \\
f_{6}=x_{3}f_{4}+(x_{4}f_{4}+f_{3}), \\
f_{5}=x_{3}(x_{4}f_{4}+f_{3}), \\
f_{4}=x_{5}f_{2}+(x_{6}f_{2}+f_{1}), \\
f_{3}=x_{5}(x_{6}f_{2}+f_{1}), \\
f_{2}=(x_{7}+x_{8}), \\
f_{1}=x_{7}x_{8}.%
\end{array}%
\end{equation}

\subsubsection{Massless case\ $(m_{1}=...=m_{8}=0)$}

\qquad This case is simple to solve loop by loop, but the advantage of the
technique NDIM is that it is possible to solve the corresponding Feynman
integral considering simultaneously all the loops. The parametric
representation of his diagram is:

\begin{equation}
G=\dfrac{(-1)^{-2D}}{\tprod\nolimits_{j=1}^{8}\Gamma (\nu _{j})}%
\dint\limits_{0}^{\infty }d\overrightarrow{x}\;\frac{\exp \left( -\dfrac{%
x_{1}f_{7}}{x_{1}f_{6}+f_{7}}p_{1}^{2}\right) }{\left(
x_{1}f_{6}+f_{7}\right) ^{\frac{D}{2}}},
\end{equation}%
where as usual the exponential is expanded, and systematically we make the
successive multiregion expansions associated to the multinomials $f_{i}$.
The order in which these multinomials appear and get expanded is:

\begin{equation}
U\longrightarrow f_{7}\longrightarrow f_{6}\longrightarrow
f_{5}\longrightarrow f_{4}\longrightarrow f_{3}\longrightarrow f_{2}.
\end{equation}%
Once the integration process finishes, finally the presolution or
multiregion expansion of the diagram $G$ is obtained:

\begin{equation}
G=\dfrac{(-1)^{-2D}}{\tprod\nolimits_{j=1}^{8}\Gamma (\nu _{j})}%
\dsum\limits_{n_{1},..,n_{15}}\phi _{n_{1},..,n_{15}}\;\dfrac{%
(p_{1}^{2})^{n_{1}}}{\Gamma (\frac{D}{2}+n_{1})}\;\Omega _{\left\{ n\right\}
}\tprod\nolimits_{j=1}^{15}\Delta _{j},
\end{equation}%
where we have defined the factor:

\begin{equation}
\Omega _{\left\{ n\right\} }=\dfrac{1}{\Gamma (-n_{1}-n_{3})\Gamma
(-n_{2}-n_{4})\Gamma (-n_{5}-n_{7})\Gamma (-n_{6}-n_{8})\Gamma
(-n_{9}-n_{11})\Gamma (-n_{10}-n_{12})},  \label{f42}
\end{equation}%
and the corresponding constraints are:

\begin{equation}
\begin{array}{lll}
\Delta _{1}=\left\langle \frac{D}{2}+n_{1}+n_{2}+n_{3}\right\rangle , &  &
\Delta _{8}=\left\langle \nu _{1}+n_{1}+n_{2}\right\rangle , \\
\Delta _{2}=\left\langle -n_{1}-n_{3}+n_{4}+n_{5}\right\rangle , &  & \Delta
_{9}=\left\langle \nu _{2}+n_{4}\right\rangle , \\
\Delta _{3}=\left\langle -n_{2}-n_{4}+n_{6}+n_{7}\right\rangle , &  & \Delta
_{10}=\left\langle \nu _{3}+n_{5}+n_{6}\right\rangle , \\
\Delta _{4}=\left\langle -n_{5}-n_{7}+n_{8}+n_{9}\right\rangle , &  & \Delta
_{11}=\left\langle \nu _{4}+n_{8}\right\rangle , \\
\Delta _{5}=\left\langle -n_{6}-n_{8}+n_{10}+n_{11}\right\rangle , &  &
\Delta _{12}=\left\langle \nu _{5}+n_{9}+n_{10}\right\rangle , \\
\Delta _{6}=\left\langle -n_{9}-n_{11}+n_{12}+n_{13}\right\rangle , &  &
\Delta _{13}=\left\langle \nu _{6}+n_{12}\right\rangle , \\
\Delta _{7}=\left\langle -n_{10}-n_{12}+n_{14}+n_{15}\right\rangle , &  &
\Delta _{14}=\left\langle \nu _{7}+n_{13}+n_{14}\right\rangle , \\
&  & \Delta _{15}=\left\langle \nu _{8}+n_{13}+n_{15}\right\rangle .%
\end{array}%
\end{equation}%
The number of possible contributions that the solution has is $C_{15}^{15}=1$%
, and it does not correspond to a series but to a single term. The solution
for this case is simply:

\begin{equation}
G=(-1)^{-2D}\dfrac{(p_{1}^{2})^{n_{1}}}{\Gamma (D/2+n_{1})}\;\Omega
_{\left\{ n\right\} }\;\dfrac{\tprod\nolimits_{j=1}^{15}\Gamma (-n_{j})}{%
\tprod\nolimits_{j=1}^{8}\Gamma (\nu _{j})},
\end{equation}%
where the indexes $n_{i}$ are to be replaced by the values:

\begin{equation}
\begin{array}{lll}
n_{1}=2D-\nu _{1}-\nu _{2}-\nu _{3}-\nu _{4}-\nu _{5}-\nu _{6}-\nu _{7}-\nu
_{8}, &  & n_{9}=D-\nu _{5}-\nu _{6}-\nu _{7}-\nu _{8}, \\
n_{2}=\nu _{2}+\nu _{3}+\nu _{4}+\nu _{5}+\nu _{6}+\nu _{7}+\nu _{8}-2D, &
& n_{10}=\nu _{6}+\nu _{7}+\nu _{8}-D, \\
n_{3}=\nu _{1}-\frac{D}{2}, &  & n_{11}=\nu _{5}-\frac{D}{2}, \\
n_{4}=-\nu _{2}, &  & n_{12}=-\nu _{6}, \\
n_{5}=\frac{3D}{2}-\nu _{3}-\nu _{4}-\nu _{5}-\nu _{6}-\nu _{7}-\nu _{8}, &
& n_{13}=\frac{D}{2}-\nu _{7}-\nu _{8}, \\
n_{6}=\nu _{4}+\nu _{5}+\nu _{6}+\nu _{7}+\nu _{8}-\frac{3D}{2}, &  &
n_{14}=\nu _{8}-\frac{D}{2}, \\
n_{7}=\nu _{3}-\frac{D}{2}, &  & n_{15}=\nu _{7}-\frac{D}{2}, \\
n_{8}=-\nu _{4}. &  &
\end{array}%
\end{equation}

\subsubsection{A massive case example$\;\left( m_{1}=...=m_{6}=0\right)
\left( m_{7}=m_{8}=m\right) $}

\begin{figure}[th]
\begin{center}
\epsfig{file=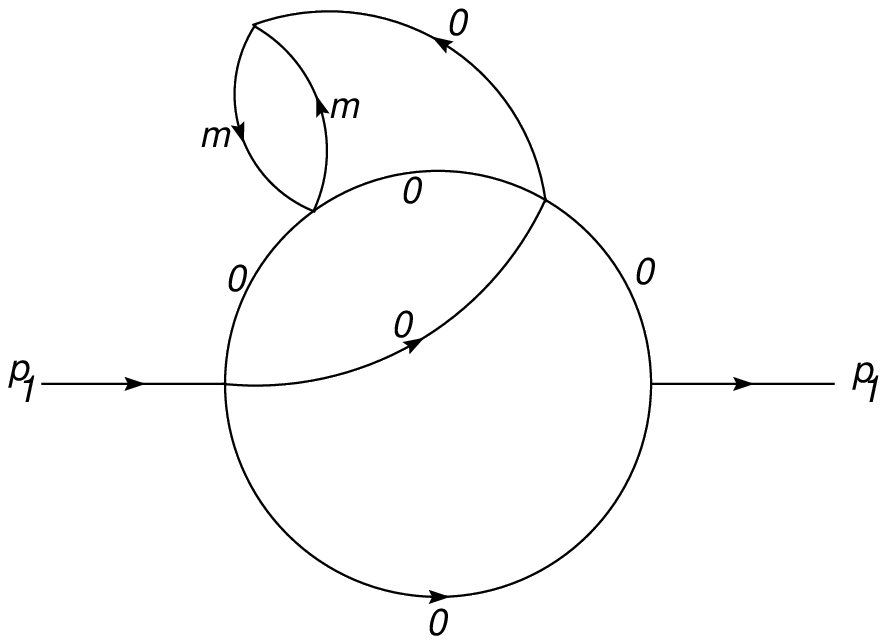,width=0.5\textwidth}
\end{center}
\caption{Mass configuration in the four-loop propagator.}
\end{figure}
Let us consider a more complicated situation, which adds one energy scale $m$%
\ to the previous problem, as it is shown in Fig. 5. Starting from the
parametric representation of the diagram:

\begin{equation}
G=\dfrac{(-1)^{-2D}}{\tprod\nolimits_{j=1}^{8}\Gamma (\nu _{j})}%
\dint\limits_{0}^{\infty }d\overrightarrow{x}\;\frac{\exp
(-f_{2}(-m^{2}))\exp \left( -\dfrac{x_{1}f_{7}}{x_{1}f_{6}+f_{7}}%
p_{1}^{2}\right) }{\left( x_{1}f_{6}+f_{7}\right) ^{\frac{D}{2}}},
\end{equation}%
and remembering that $f_{2}=(x_{7}+x_{8})$, we proceed to make the
expansions, similar to the massless case, except that this time an extra
expansion appears, due to the exponential that contains the mass scale, and
which gets expanded later on in the process or replacing the functions $%
f_{i} $. At the end of the procedure we obtain as presolution the following
expression:

\begin{equation}
G=\dfrac{(-1)^{-2D}}{\tprod\nolimits_{j=1}^{8}\Gamma (\nu _{j})}%
\dsum\limits_{n_{1},..,n_{16}}\phi _{n_{1},..,n_{16}}\;\dfrac{%
(p_{1}^{2})^{n_{1}}(-m^{2})^{n_{14}}}{\Gamma (D/2+n_{1})}\;\Omega _{\left\{
n\right\} }\tprod\nolimits_{j=1}^{15}\Delta _{j},
\end{equation}%
where $\Omega _{\left\{ n\right\} }$ is the factor:

\begin{equation}
\Omega _{\left\{ n\right\} }=\dfrac{1}{\Gamma (-n_{1}-n_{3})\Gamma
(-n_{2}-n_{4})\Gamma (-n_{5}-n_{7})\Gamma (-n_{6}-n_{8})\Gamma
(-n_{9}-n_{11})\Gamma (-n_{10}-n_{12}-n_{14})}.
\end{equation}%
The new set of constraints $\left\{ \Delta _{j}\right\} $, is given by the
following identities:

\begin{equation}
\begin{array}{lll}
\Delta _{1}=\left\langle \frac{D}{2}+n_{1}+n_{2}+n_{3}\right\rangle , &  &
\Delta _{9}=\left\langle \nu _{2}+n_{4}\right\rangle , \\
\Delta _{2}=\left\langle -n_{1}-n_{3}+n_{4}+n_{5}\right\rangle , &  & \Delta
_{10}=\left\langle \nu _{3}+n_{5}+n_{6}\right\rangle , \\
\Delta _{3}=\left\langle -n_{2}-n_{4}+n_{6}+n_{7}\right\rangle , &  & \Delta
_{11}=\left\langle \nu _{4}+n_{8}\right\rangle , \\
\Delta _{4}=\left\langle -n_{5}-n_{7}+n_{8}+n_{9}\right\rangle , &  & \Delta
_{12}=\left\langle \nu _{5}+n_{9}+n_{10}\right\rangle , \\
\Delta _{5}=\left\langle -n_{6}-n_{8}+n_{10}+n_{11}\right\rangle , &  &
\Delta _{13}=\left\langle \nu _{6}+n_{12}\right\rangle , \\
\Delta _{6}=\left\langle -n_{9}-n_{11}+n_{12}+n_{13}\right\rangle , &  &
\Delta _{14}=\left\langle \nu _{7}+n_{13}+n_{15}\right\rangle , \\
\Delta _{7}=\left\langle -n_{10}-n_{12}-n_{14}+n_{15}+n_{16}\right\rangle ,
&  & \Delta _{15}=\left\langle \nu _{8}+n_{13}+n_{16}\right\rangle , \\
\Delta _{8}=\left\langle \nu _{1}+n_{1}+n_{2}\right\rangle . &  &
\end{array}%
\end{equation}%
According to the previous analysis the number of possible contributions to
the solution of $G$ is $C_{15}^{16}=16$, of which actually only ten
contribute to the final solution. These can be distributed in the following
manner, taking into account the kinematical region of interest:

\paragraph{Solutions in the region $\left\vert \dfrac{p_{1}^{2}}{4m^{2}}%
\right\vert <1$:}

\begin{equation}
G\left( \dfrac{p_{1}^{2}}{4m^{2}}\right) =G_{1}+G_{5}+G_{9}+G_{13}.
\end{equation}

\paragraph{Solutions in the region $\left\vert \dfrac{4m^{2}}{p_{1}^{2}}%
\right\vert <1$:}

\begin{equation}
G\left( \dfrac{4m^{2}}{p_{1}^{2}}\right)
=G_{2}+G_{6}+G_{10}+G_{14}+G_{15}+G_{16}.
\end{equation}

\section{Comments}

\subsection{The factorization process}

\qquad From the point of view of the actual procedure the aspect that
optimizes the method has to do with how the multinomials that appear in
Schwinger's parametric integral are factorized. Is in this process where
resides the possibility that the integration method be generalized to $L$
loops for both massless and massive diagrams. In this respect we worked out
a procedure that allows for an adequate factorization of the multinomials $U$
and $F$ at the moment in which they are generated, which is something very
useful since without it the number of sums of these multinomials can be
quite large, depending on the number of loops and external lines that the
topology has.

In this work we have treated cases in which the multiplicity of the series
that conform the solution fulfilles the condition $\mu =\left( n-1\right) $,
where $n$ is the number of different energy scales present in the diagram.
With this criterion we have classified in three families the topologies
considered here, and for each case we give the recipe to factorize the
multinomials $U$ and $F$.

\subsubsection{Diagrams reducible recursively loop by loop}

\qquad The topologies that are included in this category are those that in
the case in which the theory does not contain masses, they can be evaluated
loop by loop by successive application of the formula corresponding to the
one-loop diagram:

\begin{equation}
G=\int \frac{d^{D}q_{1}}{i\pi ^{D/2}}\frac{1}{\left[ q_{1}^{2}\right]
^{a_{1}}\left[ (q_{1}+p_{1})^{2}\right] ^{a_{2}}},  \label{f37}
\end{equation}%
whose explicit solution is:

\begin{equation}
G=g(a_{1},a_{2})\dfrac{1}{(p^{2})^{a_{1}+a_{2}-\frac{D}{2}}},  \label{f43}
\end{equation}%
where the factor $g(a_{1},a_{2})$ is defined as:

\begin{equation}
g(a_{1},a_{2})=(-1)^{-\frac{D}{2}}\dfrac{\Gamma (a_{1}+a_{2}-\frac{D}{2}%
)\Gamma (\frac{D}{2}-a_{1})\Gamma (\frac{D}{2}-a_{2})}{\Gamma (a_{1})\Gamma
(a_{2})\Gamma (D-a_{1}-a_{2})}.
\end{equation}%
The formulae $\left( \ref{f37}\right) $ and $\left( \ref{f43}\right) $ can
be represented graphically as shown in Fig. 6.

\begin{figure}[th]
\begin{center}
\epsfig{file=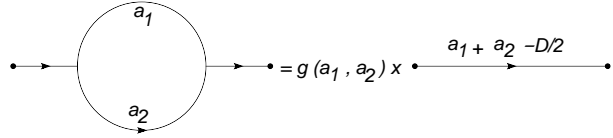,width=0.5\textwidth}
\end{center}
\caption{Graphic formula for Bubble diagram.}
\end{figure}
A situation that happens quite often together with the loop by loop
reduction is when there are two scalar propagators in series and associated
to the same mass scale, with powers $a_{1}$ and $a_{2}$ in the propagators.
In this case such powers can be summed and replaced by only one,
characterized by the power $\left( a_{1}+a_{2}\right) $. Graphically this is
represented in Fig. 7.

\begin{figure}[th]
\begin{center}
\epsfig{file=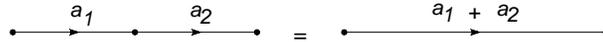,width=0.5\textwidth}
\end{center}
\caption{Reduction of propagators in serie.}
\end{figure}
The massless diagrams which applying formula $\left( \ref{f43}\right) $ can
be evaluated recursively loop by loop, are also capable of evaluation
considering all loops simultaneously and even if different mass scales are
added to the topology. That this can be done depends only on finding the
optimal factorization of $U$ and $F$. For this purpose we use the well known
topological analogy between Feynman diagram and resistive electrical
circuits, in the sense that in both 'something'\ flows in their branches and
'something'\ is conserved in their vertices. But this analogy goes further,
since electrical resistors in an electrical network are equivalent to
Schwinger%
\'{}%
s parameters in Feynman diagrams. The analogy is obvious if we observe Fig.
8.

\begin{figure}[th]
\begin{center}
\epsfig{file=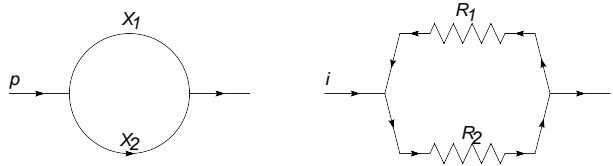,width=0.5\textwidth}
\end{center}
\caption{Simple analogy of Feynman diagram and network of resistances.}
\end{figure}
In this example the following can be seen: in the Feynman diagram one finds
that the ratio between the multinomials $F$ and $U$ is given by the
expression:

\begin{equation}
\frac{F}{U}=\frac{x_{1}x_{2}}{x_{1}+x_{2}}\Longrightarrow
F=x_{1}x_{2}\;\wedge \;U=x_{1}+x_{2}.
\end{equation}%
For simplicity we have taken $p^{2}=1$, since the results are independent of
this choice. Similarly the equivalent resistance of the electrical circuit
shown if the figure can be easily evaluated, giving:

\begin{equation}
R_{eq}=\frac{R_{1}R_{2}}{R_{1}+R_{2}}.
\end{equation}%
The resemblance of these mathematical structures and also considering the
graphic equations shown above, allow us to explain the reason why a massless
$L$-loop Feynman diagram can be reduced loop by loop in the same way that a
resistive circuit can be reduced in terms of sums of series and parallel
resistors. Both systems have the same topological characteristics.
Therefore, using this similitude it is possible to think in doing 'sums of
parameters'\ depending on whether we have series or parallel propagators,
and with this it is possible to find the ratio $\left( F/U\right) $ and thus
identify directly the multinomials $F$ and $U$. Nevertheless, the main
advantage is not really the determination of these multinomials, but the
fact that in this way it is possible to find simultaneously the adequate
factorization for the diagrams solvable directly and recursively loop by
loop, since in the process of reducing the diagram in terms of series and
parallel branches, the factorizations of the type $\left(
x_{i}+x_{j}+...+x_{k}\right) $, adequate for the application of NDIM, are
generated automatically. This is the trick used in the example $\left(
III\right) $ in order to find the factorizations. A particular case are
vacuum fluctuation diagrams, for which the multinomial $F$ vanishes, and
therefore for finding $U$ it becomes necessary to assume that the diagram
corresponds to a propagator, and for that purpose we adequately add two
external lines, in such a way that the topology be reducible loop by loop.
Then the same recipe as above is applied, the ratio $\left( F/U\right) $ is
determined and finally the already factorized multinomial $U$ is found.

Notice that the factorizations of the type $\left(
x_{i}+x_{j}+...+x_{k}\right) $ contain the parameters that belong to the
same loop (once the subtopologies of the loop propagators have been
reduced), which seems natural since the series-parallel reduction can be
done one loop at the time. It is also worth pointing out that it is possible
to do this 'parameters reduction' independently on whether the propagators
are massless or massive, which is due to the fact that the mass scales do
not appear in the multinomials $U$ and $F$ of the parametric integral.
Moreover, it is thus clear that when adding equal mass scales, these can
only be associated to the same loop, and therefore we can be sure that the
series solutions increase the multiplicity in one unit, which is logical
since one more mass scale has been added. As an example, if in the
multinomials the factor $\left( x_{i}+x_{j}+...+x_{k}\right) $ appears, it
is possible to add to the topology equal masses in various possible
situations and associate to the integral of parameters the following
exponential factors:

\begin{equation*}
\exp (x_{i}m^{2}),\;\exp (\left( x_{i}+x_{j}\right) m^{2}),\;\exp (\left(
x_{i}+x_{j}+...+x_{k}\right) m^{2}),\;etc.
\end{equation*}%
In the case in which equal masses are distributed in different loops, in
most situations it will not be possible to factorize adequately the
multinomials and the integration method will generate extra sums in the
series solutions, whose arguments would be one. The way to think about this
is that when masses are put into different loops this really corresponds to
different mass scales and therefore the multiplicity of the series solution
will grow with the different number of mass scales in the problem.

\subsubsection{Two loop topologies, with three or more external lines.}

\qquad For topologies with more than two external lines, the previous trick
is not useful. Nevertheless, there is still a generic procedure that can
provide an optimal factorization for the multinomial $U$, which in this case
is the one that determines and conditions the adequate factorization that
must be done in $F$. This procedure can be applied in general to any
two-loop diagram, although only in certain topologies optimal results are
obtained, according to the criteria expressed in $\left( \ref{f46}\right) $.
In actual practice the factorization of $U$ is done directly from the matrix
of parameters of the diagram, and since we are considering the two loop
case, $U$ is obtained from the determinant of the $2\times 2$ symmetric
submatrix that relates the loop momenta only. The general structure of the
determinant is given by the expression:

\begin{equation}
U=\left\vert
\begin{array}{cc}
X_{a}+X_{b} & \pm X_{b} \\
\pm X_{b} & X_{b}+X_{c}%
\end{array}%
\right\vert ,
\end{equation}%
where the terms $X_{a}$ and $X_{c}$ are 1-linear combinations of Schwinger
parameters, such that they do not contain common parameters among them, that
is $\left\{ X_{a}\right\} \cap \left\{ X_{c}\right\} =\varnothing $. The
term $X_{b}$ is a combination of the parameters that are common in the
diagonal elements.

The form of $U$ comes from the determinant of this matrix, that is:

\begin{equation}
U=X_{a}X_{b}+X_{b}X_{c}+X_{a}X_{c},
\end{equation}%
where we can do the following factorizations in case they are necessary:

\begin{equation}
U=\left( X_{a}+X_{c}\right) X_{b}+X_{a}X_{c},
\end{equation}

\begin{equation}
U=X_{a}X_{b}+X_{c}\left( X_{a}+X_{c}\right) ,
\end{equation}

\begin{equation}
U=X_{a}\left( X_{b}+X_{c}\right) +X_{b}X_{c}.
\end{equation}%
When adding one more mass scale, the appropriate factorization will depend
on the form in which this scale gets distributed, and them in the integrand
the following factor appears:

\begin{equation*}
\exp (X_{a}m^{2}),\;\exp (\left( X_{a}+X_{c}\right) m^{2}),\;etc.
\end{equation*}%
This provides us with a general recipe for the case of two-loop diagrams and
more than two external points. In our work only three topologies are
suitable for applying the method and producing optimal solutions, that is
they fulfill the condition that the solution multiplicity is less than the
number of energy scales of the diagram. These are indicated in Fig. 9.

\begin{figure}[ht]
\begin{center}
\epsfig{file=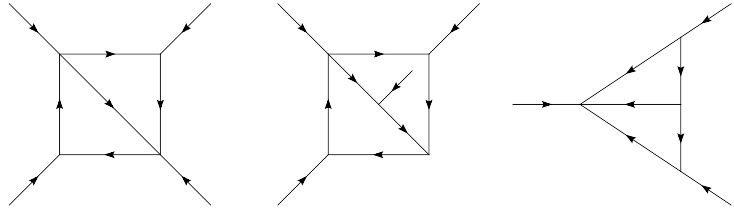,width=0.6\textwidth}
\end{center}
\caption{Two loops diagrams compatible with NDIM.}
\end{figure}

\subsubsection{The previous topologies + propagator insertions of loop by
loop reducible topologies.}

\qquad The last case we will consider is a combination of the two-loop
topologies shown in Fig. 9, with the addition of subgraphs in the loop by
loop reducible propagators, and with the advantage that it is also possible
to include masses in them. In this manner it becomes possible to extend to $%
L $-loops the two-loop topologies shown previously.

\subsection{Increase in complexity of the hypergeometric series as the
number of loops $L$ grows}

\qquad In this work we have applied the method to certain $L$-loop class of
diagrams characterized for having one, two or three different energy scales,
and such that the solutions correspond only to series of zero, one or two
variables respectively.

With this restriction in the type of solution, the topologies to which is
possible to apply the technique NDIM advantageously are the following:

\begin{itemize}
\item One loop topologies with two, three and four external points, which
have been studied in Refs. \cite{CAn2, CAn1}.

\item All $L$-loop topologies of two external points which are loop by loop
recursively reducible in the massless propagators case, but to which one or
two mass scales are added.

\item All $L$-loop vacuum fluctuation topologies that are reducible loop by
loop, and to which up to three different energy scales can be added.

\item Some two-loop diagrams, shown in Fig. 9.
\end{itemize}

\bigskip

To all these topologies more loops can be added by including one-loop
insertions in the propagators, and since it is also possible to assign mass
to these insertions, we can therefore study a great variety of graphs with
this integration technique. In conclusion a large family of $L$-loop
diagrams can be evaluated with this method.

Our results indicate that the number of parameters that characterize the
generalized hypergeometric functions that are obtained as solutions to the
diagrams increases linearly with the number of loops $L$ of the graph. In
the case of topologies that have two energy scales, the solutions are always
expressible as generalized hypergeometric series of the type $_{l}F_{l-1}$,
where $l$ corresponds to the hypergeometric order and the quantity $\left(
2l-1\right) $ is the total number of parameters that it has. Let us see how
the number of loops $L$ affects the order of the hypergeometric series in
the solution. We consider the following examples which are associated to the
diagram in Fig. 10, and which have two different energy scales $\left\{
p^{2},m^{2}\right\} $ and $L$ loops. For this class of diagrams we show two
topologies which have masses associated to one of the loops, with the
difference that while the first (A) has one massive propagator the second
(B) has both propagators massive, and the rest of the propagators are
massless.

\begin{figure}[th]
\begin{center}
\epsfig{file=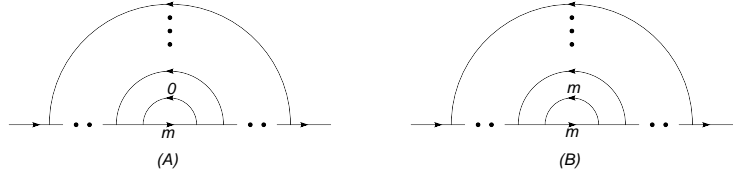,width=0.6\textwidth}
\end{center}
\caption{Two massive family of L loops propagators.}
\end{figure}
The solutions that we have found for these diagrams are given in terms of
generalized hypergeometric functions $_{l}F_{l-1}$, where the graphs with
the condition (A) have series solutions of the form:

\begin{equation*}
\;_{2L}F_{(2L-1)}\left( \left.
\begin{array}{c}
...%
\end{array}%
\right\vert
\begin{array}{c}
\frac{p^{2}}{m^{2}}%
\end{array}%
\right) ,
\end{equation*}

\begin{equation*}
\;_{2L}F_{(2L-1)}\left( \left.
\begin{array}{c}
...%
\end{array}%
\right\vert
\begin{array}{c}
\frac{m^{2}}{p^{2}}%
\end{array}%
\right) ,
\end{equation*}%
while for diagrams of the form (B) we get:

\begin{equation*}
\;_{(2L+1)}F_{2L}\left( \left.
\begin{array}{c}
...%
\end{array}%
\right\vert
\begin{array}{c}
\frac{p^{2}}{4m^{2}}%
\end{array}%
\right) ,
\end{equation*}

\begin{equation*}
\;_{(2L+1)}F_{2L}\left( \left.
\begin{array}{c}
...%
\end{array}%
\right\vert
\begin{array}{c}
\frac{4m^{2}}{p^{2}}%
\end{array}%
\right) .
\end{equation*}%
The example $\left( III\right) $ of this work corresponds precisely to a
diagram of the type (B), in which $L=4$ and whose solutions are just
hypergeometric series of type $_{9}F_{8}$. Another family of solutions that
we can consider are vacuum fluctuations of the type shown in Fig. 11,
composed of two energy scales $\left\{ m^{2},M^{2}\right\} $ and which gives
rise to solutions of the form:

\begin{equation*}
\;_{\left( 2L-2\right) }F_{(2L-3)}\left( \left.
\begin{array}{c}
...%
\end{array}%
\right\vert
\begin{array}{c}
\frac{M^{2}}{m^{2}}%
\end{array}%
\right) ,
\end{equation*}

\begin{equation*}
\;_{\left( 2L-2\right) }F_{(2L-3)}\left( \left.
\begin{array}{c}
...%
\end{array}%
\right\vert
\begin{array}{c}
\frac{m^{2}}{M^{2}}%
\end{array}%
\right) .
\end{equation*}

\begin{figure}[th]
\begin{center}
\epsfig{file=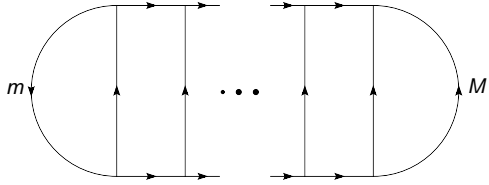,width=0.4\textwidth}
\end{center}
\caption{Two mass scales of vacuum fluctuations of L loops.}
\end{figure}
In case in which there are three energy scales, the solutions can be
expressed in terms of double hypergeometric series. In the calculations
performed in the case of one-loop diagrams \cite{CAn2}, the solutions are
completely described by Appell functions $\left(
F_{1},F_{2},F_{3},F_{4}\right) $, one Horn function $H_{2}$, and the
particular cases $S_{1}$ and $S_{2}$ of the de Kamp\'{e} de F\'{e}riet
function. Nevertheless, for $L>1$ the number of parameters that characterize
the hypergeometrics grows and then it is not possible to describe the
solutions with the same functions associated to the 1-loop results and it is
necessary to turn to a more general version of the two-variable series $F
^{\substack{ p:r:u  \\ q:s:v}}$ or $\overline{F}^{\substack{ p:r:u  \\ q:s:v
}}$.

\section{Other examples}

\qquad Although the previously shown examples are sufficient to demonstrate
the versatility and power of the method, we want to add briefly other
topologies of two, three and four external lines. The last two correspond to
specific two-loop topologies to which the method is applicable, and which
were not discussed before.

\subsection{Sunset diagram}

\begin{figure}[ht]
\begin{center}
\epsfig{file=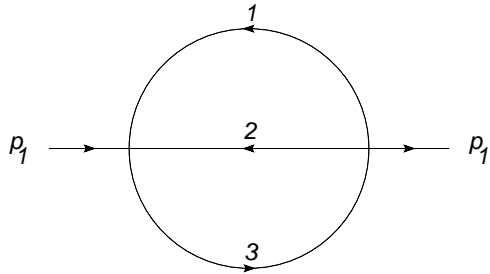,width=0.4\textwidth}
\end{center}
\caption{Sunset diagram.}
\end{figure}

\begin{equation*}
\begin{array}{ll}
U & =x_{1}x_{2}+x_{1}x_{3}+x_{2}x_{3}. \\
&  \\
F & =x_{1}x_{2}x_{3}\;p_{1}^{2}.%
\end{array}%
\end{equation*}

\subsubsection{Massless case\ $\left( m_{1}=m_{2}=m_{3}=0\right) $}

\begin{equation}
\fbox{$G=\dfrac{(-1)^{-D}}{\tprod\nolimits_{j=1}^{3}\Gamma (\nu _{j})}%
\dsum\limits_{n_{1},..,n_{4}}\phi _{n_{1},..,n_{4}}$\ $\left(
p_{1}^{2}\right) ^{n_{1}}\dfrac{\tprod\nolimits_{j=1}^{4}\Delta _{j}}{\Gamma
\left( D/2+n_{1}\right) }.$}
\end{equation}

\begin{equation}
\begin{array}{l}
\left\{
\begin{array}{l}
\Delta _{1}=\left\langle D/2+n_{1}+n_{2}+n_{3}+n_{4}\right\rangle , \\
\Delta _{2}=\left\langle \nu _{1}+n_{1}+n_{2}+n_{3}\right\rangle , \\
\Delta _{3}=\left\langle \nu _{2}+n_{1}+n_{2}+n_{4}\right\rangle , \\
\Delta _{4}=\left\langle \nu _{3}+n_{1}+n_{3}+n_{4}\right\rangle .%
\end{array}%
\right.%
\end{array}%
\end{equation}%
The solution is found immediately, and reads:

\begin{equation}
G=(-1)^{-D}\dfrac{\Gamma (D/2-\nu _{1})\Gamma (D/2-\nu _{2})\Gamma (D/2-\nu
_{3})\Gamma (\nu _{1}+\nu _{2}+\nu _{3}-D)}{\Gamma (\nu _{1})\Gamma (\nu
_{2})\Gamma (\nu _{3})\Gamma (3D/2-\nu _{1}-\nu _{2}-\nu _{3})}\left(
p_{1}^{2}\right) ^{D-\nu _{1}-\nu _{2}-\nu _{3}}.
\end{equation}

\subsubsection{Massive case I $\left( m_{1}=m_{2}=m\right) \left(
m_{3}=0\right) $}

\begin{equation}
G=\dfrac{(-1)^{-D}}{\tprod\nolimits_{j=1}^{3}\Gamma (\nu _{j})}%
\dint\limits_{0}^{\infty }d\overrightarrow{x}\;\frac{\exp \left( -f\left(
-m^{2}\right) \right) \exp \left( -F/U\right) }{U^{D/2}}.
\end{equation}%
For this particular case $U$ has to be factorized as:

\begin{equation}
\begin{array}{l}
U=x_{1}x_{2}+x_{3}f, \\
\\
f=\left( x_{1}+x_{2}\right) .%
\end{array}%
\end{equation}

\begin{equation}
\fbox{$G=\dfrac{(-1)^{-D}}{\tprod\nolimits_{j=1}^{3}\Gamma (\nu _{j})}%
\dsum\limits_{n_{1},..,n_{6}}\phi _{n_{1},..,n_{6}}$\ $\left( -m^{2}\right)
^{n_{1}}\left( p_{1}^{2}\right) ^{n_{2}}\dfrac{\tprod\nolimits_{j=1}^{5}%
\Delta _{j}}{\Gamma \left( D/2+n_{2}\right) \Gamma (-n_{2}-n_{4})}.$}
\end{equation}

\begin{equation}
\begin{array}{l}
\left\{
\begin{array}{l}
\Delta _{1}=\left\langle D/2+n_{2}+n_{3}+n_{4}\right\rangle , \\
\Delta _{2}=\left\langle -n_{1}-n_{4}+n_{5}+n_{6}\right\rangle , \\
\Delta _{3}=\left\langle \nu _{1}+n_{2}+n_{3}+n_{5}\right\rangle , \\
\Delta _{4}=\left\langle \nu _{2}+n_{2}+n_{3}+n_{6}\right\rangle , \\
\Delta _{5}=\left\langle \nu _{3}+n_{2}+n_{4}\right\rangle .%
\end{array}%
\right.%
\end{array}%
\end{equation}%
The solutions, in terms of the contributions $G_{j}$, are:

\begin{equation}
\begin{array}{l}
G\left( \dfrac{m^{2}}{p_{1}^{2}}\right) =G_{1}+G_{4}+G_{5}+G_{6}, \\
\\
G\left( \dfrac{p_{1}^{2}}{m^{2}}\right) =G_{2}.%
\end{array}%
\end{equation}

\subsubsection{Massive case II\ $\left( m_{1}=m_{2}=m\right) \left(
m_{3}=M\right) $}

\begin{equation}
G=\dfrac{(-1)^{-D}}{\tprod\nolimits_{j=1}^{3}\Gamma (\nu _{j})}%
\dint\limits_{0}^{\infty }d\overrightarrow{x}\;\frac{\exp \left( -f\left(
-m^{2}\right) -x_{3}\left( -M^{2}\right) \right) \exp \left( -F/U\right) }{%
U^{D/2}}.
\end{equation}

\begin{equation}
\begin{array}{l}
U=x_{1}x_{2}+x_{3}f, \\
\\
f=\left( x_{1}+x_{2}\right) .%
\end{array}%
\end{equation}

\begin{equation}
\fbox{$G=\dfrac{(-1)^{-D}}{\tprod\nolimits_{j=1}^{3}\Gamma (\nu _{j})}%
\dsum\limits_{n_{1},..,n_{7}}\phi _{n_{1},..,n_{7}}$\ $\left( -m^{2}\right)
^{n_{1}}\left( -M^{2}\right) ^{n_{2}}\left( p_{1}^{2}\right) ^{n_{3}}\dfrac{%
\tprod\nolimits_{j=1}^{5}\Delta _{j}}{\Gamma \left( D/2+n_{3}\right) \Gamma
(-n_{1}-n_{5})}.$}
\end{equation}

\begin{equation}
\begin{array}{l}
\left\{
\begin{array}{l}
\Delta _{1}=\left\langle D/2+n_{3}+n_{4}+n_{5}\right\rangle , \\
\Delta _{2}=\left\langle -n_{1}-n_{5}+n_{6}+n_{7}\right\rangle , \\
\Delta _{3}=\left\langle \nu _{1}+n_{3}+n_{4}+n_{6}\right\rangle , \\
\Delta _{4}=\left\langle \nu _{2}+n_{3}+n_{4}+n_{7}\right\rangle , \\
\Delta _{5}=\left\langle \nu _{3}+n_{2}+n_{3}+n_{5}\right\rangle .%
\end{array}%
\right.%
\end{array}%
\end{equation}%
The solutions determine three kinematical regions, which we write in terms
of the contributions $G_{i,j}$:

\begin{equation}
\begin{array}{l}
G\left( \dfrac{4m^{2}}{p_{1}^{2}},\dfrac{M^{2}}{p_{1}^{2}}\right)
=G_{1,2}+G_{1,4}+G_{2,5}+G_{2,6}+G_{2,7}+\left( G_{4,5}=0\right)
+G_{4,6}+G_{4,7}\Longrightarrow F^{\substack{ 2:2:0  \\ 0:3:1}}\left(
...\right) \\
\\
G\left( \dfrac{M^{2}}{4m^{2}},\dfrac{p_{1}^{2}}{4m^{2}}\right)
=G_{2,3}+G_{3,4}\Longrightarrow F^{\substack{ 4:0:0  \\ 2:1:1}}\left(
...\right) \\
\\
G\left( \dfrac{4m^{2}}{M^{2}},\dfrac{p_{1}^{2}}{M^{2}}\right)
=G_{1,3}+G_{3,5}+G_{3,6}+G_{3,7}\Longrightarrow F^{\substack{ 2:2:0  \\ %
0:3:1 }}\left( ...\right)%
\end{array}%
\end{equation}%
In each of these cases the contributions $G_{i,j}$ correspond to de Kamp\'{e}
de F\'{e}riet functions $(F^{\substack{ p:r:u  \\ q:s:v}})$.

\subsection{Massless non-planar Double-Box diagram (On-Shell case)}

\begin{figure}[th]
\begin{center}
\epsfig{file=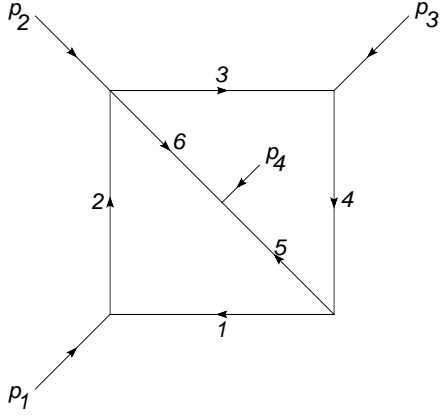,width=0.36\textwidth}
\end{center}
\caption{Labelled non-planar Box diagram.}
\end{figure}

\begin{equation}
\begin{array}{ll}
U & =\left( x_{1}+x_{2}\right) \left( x_{5}+x_{6}\right) +\left(
x_{3}+x_{4}\right) \left( x_{5}+x_{6}\right) +\left( x_{1}+x_{2}\right)
\left( x_{3}+x_{4}\right) , \\
&  \\
F & =x_{1}x_{3}x_{6}\;s+x_{2}x_{3}x_{5}\;u+x_{2}x_{4}x_{6}\;t,%
\end{array}%
\end{equation}

\begin{equation}
\fbox{$G=A\dsum\limits_{n_{1},..,n_{12}}\phi _{n_{1},..,n_{12}}$\ $\dfrac{%
\left( s\right) ^{n_{1}}\left( u\right) ^{n_{2}}\left( t\right) ^{n_{3}}}{%
\Gamma (D/2+n_{1}+n_{2}+n_{3})}\dfrac{\tprod\nolimits_{j=1}^{10}\Delta _{j}}{%
\Gamma (-n_{4}-n_{6})\Gamma (-n_{5}-n_{6})\Gamma (-n_{4}-n_{5})}.$}
\end{equation}

\begin{equation*}
A=\dfrac{(-1)^{-D}}{\tprod\nolimits_{j=1}^{6}\Gamma (\nu _{j})},
\end{equation*}

\begin{equation*}
\begin{array}{l}
\left\{
\begin{array}{l}
\begin{array}{l}
\Delta _{1}=\left\langle
D/2+n_{1}+n_{2}+n_{3}+n_{4}+n_{5}+n_{6}\right\rangle , \\
\Delta _{2}=\left\langle -n_{4}-n_{6}+n_{7}+n_{8}\right\rangle , \\
\Delta _{3}=\left\langle -n_{5}-n_{6}+n_{9}+n_{10}\right\rangle , \\
\Delta _{4}=\left\langle -n_{4}-n_{5}+n_{11}+n_{12}\right\rangle , \\
\Delta _{5}=\left\langle \nu _{1}+n_{1}+n_{7}\right\rangle , \\
\Delta _{6}=\left\langle \nu _{2}+n_{2}+n_{3}+n_{8}\right\rangle , \\
\Delta _{7}=\left\langle \nu _{3}+n_{1}+n_{2}+n_{9}\right\rangle , \\
\Delta _{8}=\left\langle \nu _{4}+n_{3}+n_{10}\right\rangle , \\
\Delta _{9}=\left\langle \nu _{5}+n_{2}+n_{11}\right\rangle , \\
\Delta _{10}=\left\langle \nu _{6}+n_{1}+n_{3}+n_{12}\right\rangle .%
\end{array}%
\end{array}%
\right.%
\end{array}%
\end{equation*}%
The sets of terms which contribute to the solution are shown as functions of
the components $G_{i,j}$ and of the type of two-valued function that
represents them:

\bigskip

\begin{itemize}
\item[Set 1 :] $G_{1,2}+G_{1,12}+G_{2,8}+G_{8,12}\Longrightarrow F
^{\substack{ 2:1:1  \\ 1:1:1}}\left( \left.
\begin{array}{cc}
&
\end{array}%
\right\vert -\dfrac{u}{t},-\dfrac{s}{t}\right) $

\item[Set 2 :] $G_{2,3}+G_{2,9}+G_{3,12}+G_{9,12}\Longrightarrow F
^{\substack{ 2:1:1  \\ 1:1:1}}\left( \left.
\begin{array}{cc}
&
\end{array}%
\right\vert -\dfrac{u}{s},-\dfrac{t}{s}\right) $

\item[Set 3 :] $G_{1,3}+G_{1,9}+G_{3,8}+G_{8,9}\Longrightarrow F^{\substack{ %
2:1:1  \\ 1:1:1}}\left( \left.
\begin{array}{cc}
&
\end{array}%
\right\vert -\dfrac{s}{u},-\dfrac{t}{u}\right) $

\item[Set 4 :] $G_{1,11}+G_{8,11}\Longrightarrow \overline{F}^{\substack{ %
2:1:2  \\ 1:1:0}}\left( \left.
\begin{array}{cc}
&
\end{array}%
\right\vert -\dfrac{s}{t},\dfrac{t}{u}\right) $

\item[Set 5 :] $G_{3,11}+G_{9,11}\Longrightarrow \overline{F}^{\substack{ %
2:1:2  \\ 1:1:0}}\left( \left.
\begin{array}{cc}
&
\end{array}%
\right\vert -\dfrac{t}{s},\dfrac{s}{u}\right) $

\item[Set 6:] $G_{1,10}+G_{8,10}\Longrightarrow \overline{F}^{\substack{ %
2:1:2  \\ 1:1:0}}\left( \left.
\begin{array}{cc}
&
\end{array}%
\right\vert -\dfrac{s}{u},\dfrac{u}{t}\right) $

\item[Set 7 :] $G_{2,7}\Longrightarrow \overline{F}^{\substack{ 2:1:2  \\ %
1:1:0}}\left( \left.
\begin{array}{cc}
&
\end{array}%
\right\vert -\dfrac{u}{t},\dfrac{t}{s}\right) $

\item[Set 8 :] $G_{7,12}\Longrightarrow \overline{F}^{\substack{ 1:2:1  \\ %
2:0:1}}\left( \left.
\begin{array}{cc}
&
\end{array}%
\right\vert -\dfrac{t}{s},\dfrac{u}{t}\right) $

\item[Set 9 :] $G_{2,10}\Longrightarrow \overline{F}^{\substack{ 2:1:2  \\ %
1:1:0}}\left( \left.
\begin{array}{cc}
&
\end{array}%
\right\vert -\dfrac{u}{s},\dfrac{s}{t}\right) $

\item[Set 10 :] $G_{10,12}\Longrightarrow \overline{F}^{\substack{ 1:2:1  \\ %
2:0:1}}\left( \left.
\begin{array}{cc}
&
\end{array}%
\right\vert -\dfrac{s}{t},\dfrac{u}{s}\right) $

\item[Set 11 :] $G_{3,7}\Longrightarrow \overline{F}^{\substack{ 2:1:2  \\ %
1:1:0}}\left( \left.
\begin{array}{cc}
&
\end{array}%
\right\vert -\dfrac{t}{u},\dfrac{u}{s}\right) $

\item[Set 12 :] $G_{7,9}\Longrightarrow \overline{F}^{\substack{ 1:2:1  \\ %
2:0:1}}\left( \left.
\begin{array}{cc}
&
\end{array}%
\right\vert -\dfrac{u}{s},\dfrac{t}{u}\right) $

\item[Set 13 :] $G_{7,10}\Longrightarrow F^{\substack{ 1:2:2  \\ 2:0:0}}%
\left( \left.
\begin{array}{cc}
&
\end{array}%
\right\vert -\dfrac{u}{t},-\dfrac{u}{s}\right) $

\item[Set 14 :] $G_{10,11}\Longrightarrow F^{\substack{ 1:2:2  \\ 2:0:0}}%
\left( \left.
\begin{array}{cc}
&
\end{array}%
\right\vert -\dfrac{s}{t},-\dfrac{s}{u}\right) $

\item[Set 15 :] $G_{7,11}\Longrightarrow F^{\substack{ 1:2:2  \\ 2:0:0}}%
\left( \left.
\begin{array}{cc}
&
\end{array}%
\right\vert -\dfrac{t}{s},-\dfrac{t}{u}\right) .$
\end{itemize}

\bigskip

This example has been solved with the usual NDIM method in Ref. \cite{ASu10}%
. Nevertheless, it is evident that doing it the way presented here the
solutions are found more directly. They are of the type that corresponds to
three different energy scales, that is two-variable series, although without
generating a big number of contributions to the solution, neither redundant
sums with unit argument in each one of them. Table III allows to visualize
more precisely the improvement of our optimization, comparing the solutions
obtained without any simplification and with the present method.

\begin{equation}
\begin{tabular}{lll}
\hline
& Previous NDIM & New NDIM \\ \hline
Possible solutions & \multicolumn{1}{c}{6435} & \multicolumn{1}{c}{66} \\
Relevant solutions & \multicolumn{1}{c}{2916} & \multicolumn{1}{c}{27} \\
Multiplicity$\text{ }\left( {\mu }\right) $ & \multicolumn{1}{c}{8} &
\multicolumn{1}{c}{2} \\ \hline
\end{tabular}
\tag{Table III}
\end{equation}

\subsection{Four-loop vertex}

\begin{figure}[th]
\begin{center}
\epsfig{file=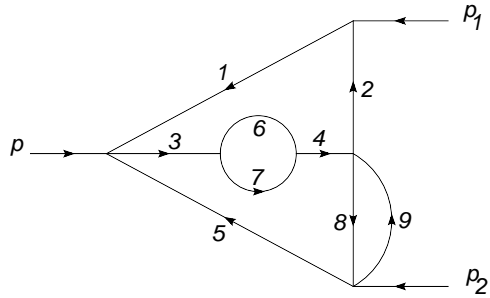,width=0.4\textwidth}
\end{center}
\caption{Labelled four-loop vertex diagram.}
\end{figure}

\subsubsection{A case with three energy scales $\left(
p_{1}^{2}=p_{2}^{2}=0\right) $ and $\left( m_{6}=m\text{, }m_{9}=M\right) $}

\begin{equation}
G=\dfrac{(-1)^{-2D}}{\tprod\nolimits_{j=1}^{9}\Gamma (\nu _{j})}%
\dint\limits_{0}^{\infty }d\overrightarrow{x}\;\frac{\exp \left(
-x_{6}\left( -m^{2}\right) \right) \exp \left( -x_{9}\left( -M^{2}\right)
\right) \exp \left( -F/U\right) }{U^{D/2}},
\end{equation}

\begin{equation}
\begin{array}{l}
F=x_{1}x_{5}f_{2}f_{5}\;p^{2}, \\
\\
U=f_{1}f_{4}f_{6}+f_{1}f_{2}f_{5}+f_{5}f_{6},%
\end{array}%
\end{equation}%
where the functions $f_{i}$ are given by:

\begin{equation}
\begin{array}{l}
f_{6}=x_{8}x_{9}+x_{5}f_{2}, \\
f_{5}=x_{6}x_{7}+f_{3}f_{4}, \\
f_{4}=x_{6}+x_{7}, \\
f_{3}=x_{3}+x_{4}, \\
f_{2}=x_{8}+x_{9}, \\
f_{1}=x_{1}+x_{2}.%
\end{array}%
\end{equation}

\begin{equation}
\fbox{$G=\dfrac{(-1)^{-2D}}{\tprod\nolimits_{j=1}^{9}\Gamma (\nu _{j})}%
\dsum\limits_{n_{1},..,n_{18}}\phi _{n_{1},..,n_{18}}$\ $\dfrac{%
(-m^{2})^{n_{1}}(-M^{2})^{n_{2}}(p^{2})^{n_{3}}}{\Gamma (\frac{D}{2}+n_{3})}$%
\ $\Omega _{\left\{ n\right\} }\tprod\nolimits_{j=1}^{16}\Delta _{j}$},
\end{equation}%
where the factor $\Omega _{\left\{ n\right\} }$ is:

\begin{equation}
\Omega _{\left\{ n\right\} }=\dfrac{1}{\Gamma (-n_{4}-n_{5})\Gamma
(-n_{3}-n_{5}-n_{6})\Gamma (-n_{4}-n_{6})\Gamma (-n_{3}-n_{5}-n_{12})\Gamma
(-n_{10})\Gamma (-n_{4}-n_{10})},
\end{equation}

\begin{equation*}
\begin{array}{l}
\left\{
\begin{array}{l}
\begin{array}{l}
\Delta _{1}=\left\langle D/2+n_{3}+n_{4}+n_{5}+n_{6}\right\rangle , \\
\Delta _{2}=\left\langle -n_{4}-n_{5}+n_{7}+n_{8}\right\rangle , \\
\Delta _{3}=\left\langle -n_{3}-n_{5}-n_{6}+n_{9}+n_{10}\right\rangle , \\
\Delta _{4}=\left\langle -n_{4}-n_{6}+n_{11}+n_{12}\right\rangle , \\
\Delta _{5}=\left\langle -n_{3}-n_{5}-n_{12}+n_{13}+n_{14}\right\rangle , \\
\Delta _{6}=\left\langle -n_{10}+n_{15}+n_{16}\right\rangle , \\
\Delta _{7}=\left\langle -n_{4}-n_{10}+n_{17}+n_{18}\right\rangle , \\
\Delta _{8}=\left\langle \nu _{1}+n_{3}+n_{7}\right\rangle ,%
\end{array}%
\end{array}%
\begin{array}{ll}
& \Delta _{9}=\left\langle \nu _{2}+n_{8}\right\rangle , \\
& \Delta _{10}=\left\langle \nu _{3}+n_{15}\right\rangle , \\
& \Delta _{11}=\left\langle \nu _{4}+n_{16}\right\rangle , \\
& \Delta _{12}=\left\langle \nu _{5}+n_{3}+n_{12}\right\rangle , \\
& \Delta _{13}=\left\langle \nu _{6}+n_{1}+n_{9}+n_{17}\right\rangle , \\
& \Delta _{14}=\left\langle \nu _{7}+n_{9}+n_{18}\right\rangle , \\
& \Delta _{15}=\left\langle \nu _{8}+n_{11}+n_{13}\right\rangle , \\
& \Delta _{16}=\left\langle \nu _{9}+n_{2}+n_{11}+n_{14}\right\rangle .%
\end{array}%
\right.%
\end{array}%
\end{equation*}

The sets of solutions are given by the following equations in terms of the
contributions $G_{i,j}$.

\bigskip

\begin{itemize}
\item[Set 1 :] $G_{1,2}+G_{1,5}+G_{1,13}+G_{2,4}+G_{2,18}+\left(
G_{4,5}=0\right) +G_{4,13}+G_{5,18}+G_{13,18}\Longrightarrow F^{\substack{ %
3:3:2  \\ 2:3:2}}\left( \left.
\begin{array}{cc}
&
\end{array}%
\right\vert \dfrac{m^{2}}{p^{2}},-\dfrac{M^{2}}{p^{2}}\right) $

\item[Set 2 :] $G_{1,3}+G_{3,4}+G_{3,18}\Longrightarrow F^{\substack{ 3:3:2
\\ 2:3:2}}\left( \left.
\begin{array}{cc}
&
\end{array}%
\right\vert \dfrac{m^{2}}{M^{2}},-\dfrac{p^{2}}{M^{2}}\right) $

\item[Set 3 :] $G_{2,3}+G_{3,5}+G_{3,13}\Longrightarrow F^{\substack{ 4:2:2
\\ 3:2:2}}\left( \left.
\begin{array}{cc}
&
\end{array}%
\right\vert \dfrac{M^{2}}{m^{2}},\dfrac{p^{2}}{m^{2}}\right) $

\item[Set 4 :] $G_{7,9}+G_{9,12}\Longrightarrow F^{\substack{ 2:4:3  \\ %
3:2:1 }}\left( \left.
\begin{array}{cc}
&
\end{array}%
\right\vert \dfrac{M^{2}}{m^{2}},-\dfrac{M^{2}}{p^{2}}\right) $

\item[Set 5 :] $G_{7,11}+\left( G_{11,12}=0\right) \Longrightarrow F
^{\substack{ 3:3:3  \\ 4:1:1}}\left( \left.
\begin{array}{cc}
&
\end{array}%
\right\vert \dfrac{m^{2}}{M^{2}},\dfrac{m^{2}}{p^{2}}\right) $

\item[Set 6 :] $G_{9,11}\Longrightarrow F^{\substack{ 2:4:3  \\ 3:2:1}}%
\left( \left.
\begin{array}{cc}
&
\end{array}%
\right\vert \dfrac{p^{2}}{m^{2}},-\dfrac{p^{2}}{M^{2}}\right) $

\item[Set 7 :] $G_{1,11}+G_{4,11}\Longrightarrow \overline{F}^{\substack{ %
3:3:3  \\ 2:3:1}}\left( \left.
\begin{array}{cc}
&
\end{array}%
\right\vert \dfrac{m^{2}}{p^{2}},\dfrac{p^{2}}{M^{2}}\right) $

\item[Set 8 :] $G_{11,18}\Longrightarrow \overline{F}^{\substack{ 2:3:3  \\ %
3:1:3}}\left( \left.
\begin{array}{cc}
&
\end{array}%
\right\vert -\dfrac{p^{2}}{M^{2}},-\dfrac{m^{2}}{p^{2}}\right) $

\item[Set 9 :] $G_{2,9}+G_{5,9}\Longrightarrow \overline{F}^{\substack{ %
3:2:4  \\ 2:2:2}}\left( \left.
\begin{array}{cc}
&
\end{array}%
\right\vert -\dfrac{M^{2}}{p^{2}},-\dfrac{p^{2}}{m^{2}}\right) $

\item[Set 10 :] $G_{9,13}\Longrightarrow \overline{F}^{\substack{ 2:4:2  \\ %
3:2:2}}\left( \left.
\begin{array}{cc}
&
\end{array}%
\right\vert \dfrac{p^{2}}{m^{2}},\dfrac{M^{2}}{p^{2}}\right) $

\item[Set 11 :] $G_{1,7}+G_{1,12}+G_{4,7}+G_{4,12}\Longrightarrow \overline{F%
}^{\substack{ 3:3:3  \\ 2:3:1}}\left( \left.
\begin{array}{cc}
&
\end{array}%
\right\vert \dfrac{m^{2}}{M^{2}},\dfrac{M^{2}}{p^{2}}\right) $

\item[Set 12 :] $G_{7,18}+G_{12,18}\Longrightarrow \overline{F}^{\substack{ %
2:3:3  \\ 3:1:3}}\left( \left.
\begin{array}{cc}
&
\end{array}%
\right\vert -\dfrac{M^{2}}{p^{2}},-\dfrac{m^{2}}{M^{2}}\right) $

\item[Set 13 :] $G_{3,9}\Longrightarrow \overline{F}^{\substack{ 3:2:4  \\ %
2:2:2}}\left( \left.
\begin{array}{cc}
&
\end{array}%
\right\vert -\dfrac{p^{2}}{M^{2}},-\dfrac{M^{2}}{m^{2}}\right) $

\item[Set 14 :] $G_{2,7}+G_{2,12}+G_{5,7}+G_{5,12}\Longrightarrow \overline{F%
}^{\substack{ 4:2:3  \\ 3:2:1}}\left( \left.
\begin{array}{cc}
&
\end{array}%
\right\vert \dfrac{M^{2}}{m^{2}},-\dfrac{m^{2}}{p^{2}}\right) $

\item[Set 15 :] $G_{7,13}+G_{12,13}\Longrightarrow \overline{F}^{\substack{ %
3:3:2  \\ 4:1:2}}\left( \left.
\begin{array}{cc}
&
\end{array}%
\right\vert \dfrac{m^{2}}{p^{2}},-\dfrac{M^{2}}{m^{2}}\right) $

\item[Set 16 :] $G_{3,11}\Longrightarrow \overline{F}^{\substack{ 4:2:3  \\ %
3:2:1}}\left( \left.
\begin{array}{cc}
&
\end{array}%
\right\vert \dfrac{p^{2}}{m^{2}},-\dfrac{m^{2}}{M^{2}}\right) .$
\end{itemize}

In this example the fact that $\left( M^{2}>m^{2}\right) $ or viceversa
leaves out automatically sets of solutions, since only one of these
conditions can appear in the same topology.

\section{Conclusions}

\qquad The integration method here presented is a powerful tool for solving
Feynman integrals with arbitrary exponents in the propagators. Its main
characteristics resides in its simplicity in finding the solution of
diagrams, even in the evaluation of $L$-loop diagrams. This technique, which
can be more appropriately called Integration by Fractional Expansion, is
clearly an important competitor to other Feynman diagram evaluation methods.
In spite of its generality, we have found that the technique is adequately
applicable to certain types of graphs, obtaining solutions which we call
optimal, in the sense that they are algebraically manageable, and whose
complexity does not depend on the number of loops $L$ of the diagram.
Although this method can be applied to any diagram, the complexity of the
solutions in the more general cases has a high dependence on $L$, which
implies a large number of terms as well as a high multiplicity $\mu $ of the
resulting series expansions, which makes the analysis of the solution highly
non trivial. Imposing to evaluate only topologies whose solutions can be
represented as series in which the variables are expressed as ratios between
the different energy scales of the diagram, determines that the multiplicity
of each of them fulfills the relation $\mu =\left( n-1\right) $, where $n$
is the number of different energy scales in the diagram. This is the
restriction that allowed us to find that the method here proposed is
applicable in a natural way to certain $L$-loop class of diagrams, which in
the massless case happen to be solvable loop by loop. Using this condition
it becomes possible, depending on the number of energy scales present in
this class of topologies, obtain solutions as expansions in multivalued
series. Nevertheless, we have considered adequate to evaluate topologies
whose solutions could be expressed at most in terms of two-variable series,
because hypergeometric series of this type have been amply treated in the
literature. The advantages of this integration method can be summarized in
the following:

\begin{itemize}
\item For the family of $L$ loop graphs, reducible recursively loop by loop,
the application of the technique makes possible to find the solution
considering all loops simultaneously, through its Schwinger parametric
representation.

\item The integration method allows to extend in a simple manner the
solutions to cases in which different mass scales are considered in the
graph.

\item It is possible to apply the integration method to certain three and
four external line topologies, and generalize it to $L$ loops making
one-loop insertions in the propagators. In general the application of this
technique can be extended to more external lines, but the solutions are
series with $\mu >2$, and therefore have not been considered in this work.

\item Independently of the number of loops, in topologies of the loop by
loop reducible type, the multiplicity $\mu $ of the hypergeometric series
solution does not have any dependence on the number of loops $L$, and only
depends on the number of different energy scales that characterize the
graph. For topologies that are not in the class reducible loop by loop,
except for the special cases described previously, the multiplicity grows
when $L$ increases. This fact makes this type of topologies not optimal for
the application of the method, because the complexity of the solutions grows
very quickly with $L$. We can estimate this growth in terms of the
contributions that the solution contains, using the following expression:
\end{itemize}

\bigskip

\begin{equation*}
\text{Number of contributions of the solution}\sim \frac{\left( \mu
(L)+\delta \right) !}{\delta !\left( \mu (L)\right) !}.
\end{equation*}

\bigskip

A significant fact is that the method allows to easily add masses to any
type of topologies. In general for generic diagrams we have been able to
conclude that:

\begin{itemize}
\item If we add $M$ different mass scales, the multiplicity of the series
expressions that are part of the solution grows in $M$ units.

\item In most cases if we add masses to propagators belonging to $l$
different loops, the masses have to be considered different, even if they
have the same value, and the multiplicity increases in $l$ units.

\item The conjunction of the previous premises implies that if one
associates $M$ equal masses to the propagators of a loop, the multiplicity
of the series solution only increases in one unit.
\end{itemize}

\newpage

\appendix

\section{Mathematical formalism of NDIM}

\qquad In section $\left( 2.2\right) $ the technique was introduced, and
also its origins starting from the identity associated to the loop integral
parametrization and known as Schwinger's parametrization:

\begin{equation}
\frac{1}{A^{\beta }}=\frac{1}{\Gamma (\beta )}\dint\limits_{0}^{\infty
}dx\;x^{\beta -1}\exp (-Ax),  \label{f1}
\end{equation}%
and furthermore the equivalence between the integral symbol and a delta
Kronecker $\left( \ref{f28}\right)$ was shown starting from the equation $%
\left( \ref{f1}\right)$:

\begin{equation}
\dint dx\;x^{\beta +n-1}\equiv \Gamma \left( \beta \right) \dfrac{\Gamma
\left( n+1\right) }{\left( -1\right) ^{n}}\;\delta _{\beta +n,0}.  \label{f2}
\end{equation}%
For simplicity we have eliminated the limits of the integral, since this
identity has validity only in the context of the expansion of the integrand
in $\left( \ref{f1}\right) $. This expression is crucial for the development
of the method NDIM, since in the evaluation of Feynman diagrams the
corresponding Schwinger parametric representation is a generalized structure
of the expression $\left( \ref{f1}\right) $.

\subsection{Some properties}

\qquad In order to study some of the properties of $\left( \ref{f2}\right) $%
, it is convenient to introduce a useful notation for helping us to
formalize the mechanism of the technique NDIM. Thus:

\begin{equation}
\int dx\;x^{\nu _{1}+\nu _{2}-1}\equiv \left\langle \nu _{1}+\nu
_{2}\right\rangle ,  \label{f4}
\end{equation}%
where $\nu _{1}$ and $\nu _{2}$ are indexes that can assume arbitrary values.

\subsubsection{Property l. Commutativity of indexes}

\qquad We can write $\left( \ref{f4}\right) $ explicitly in two possible
forms, according to formula $\left( \ref{f2}\right) $, and then:

\begin{equation}
\begin{array}{cc}
\left\langle \nu _{1}+\nu _{2}\right\rangle = & \left\{
\begin{array}{c}
\Gamma \left( \nu _{1}\right) \dfrac{\Gamma \left( \nu _{2}+1\right) }{%
\left( -1\right) ^{\nu _{2}}}\;\delta _{\nu _{1}+\nu _{2},0}, \\
\\
\Gamma \left( \nu _{2}\right) \dfrac{\Gamma \left( \nu _{1}+1\right) }{%
\left( -1\right) ^{\nu _{1}}}\;\delta _{\nu _{1}+\nu _{2},0}.%
\end{array}%
\right.%
\end{array}
\label{f10}
\end{equation}%
Starting from $\left( \ref{f1}\right) $ it is possible to show the
equivalence of both forms in $\left( \ref{f10}\right) $, and for this
purpose it is enough to expand the exponential of the integrand and replace $%
\left\langle \cdot \right\rangle $ for this case.

Let us consider the following integral representation:

\begin{equation}
\frac{1}{A^{\nu _{1}}}=\frac{1}{\Gamma \left( \nu _{1}\right) }%
\dint\limits_{0}^{\infty }dx\;x^{\nu _{1}-1}\exp \left( -Ax\right) ,
\label{f5}
\end{equation}%
which in terms of series, in the sense proposed $\left( \ref{f4}\right) $,
turns out to be:

\begin{equation}
\frac{1}{A^{\nu _{1}}}=\frac{1}{\Gamma \left( \nu _{1}\right) }%
\dsum\limits_{\nu _{2}}\frac{\left( -1\right) ^{\nu _{2}}}{\Gamma \left( \nu
_{2}+1\right) }A^{\nu _{2}}\left\langle \nu _{1}+\nu _{2}\right\rangle .
\end{equation}%
Selecting now $\left\langle \nu _{1}+\nu _{2}\right\rangle =\dfrac{\Gamma
\left( \nu _{1}\right) \Gamma \left( \nu _{2}+1\right) }{\left( -1\right)
^{\nu _{2}}}\;\delta _{\nu _{1}+\nu _{2},0}$, we directly obtain the
equality in $\left( \ref{f5}\right) $. Analogously selecting $\left\langle
\nu _{1}+\nu _{2}\right\rangle =\dfrac{\Gamma \left( \nu _{2}\right) \Gamma
\left( \nu _{1}+1\right) }{\left( -1\right) ^{\nu _{1}}}\;\delta _{\nu
_{1}+\nu _{2},0}$ and using afterwards the identity:

\begin{equation}
\dfrac{\Gamma \left( y\right) }{\Gamma \left( y-z\right) }=\left( -1\right)
^{-z}\dfrac{\Gamma \left( 1+z-y\right) }{\Gamma \left( 1-y\right) },
\end{equation}%
with $y=\nu _{1}$ and $z=2\nu _{1}$, the equality $\left( \ref{f5}\right) $
is obtained once again, and therefore the equivalence between both forms of
writing $\left\langle \nu _{1}+\nu _{2}\right\rangle $ is quickly
established.

Another form of writing $\left( \ref{f4}\right) $, which is quite useful for
simplifying terms that contain the factor:

\begin{equation}
\frac{1}{\Gamma (m+1)},
\end{equation}%
where $m$ is an arbitrary index, is the following:

\begin{equation}
\left\langle \nu _{1}+\nu _{2}\right\rangle =\left\langle \nu _{1}+\nu
_{2}-m+m\right\rangle =\left\langle -m+m\right\rangle =\Gamma (-m)\frac{%
\Gamma (m+1)}{(-1)^{m}}\;\delta _{\nu _{1}+\nu _{2},0}.
\end{equation}%
Notice that we have made explicit use of the Kronecker delta in order to
simplify the parenthesis $\left\langle \cdot \right\rangle $, and since this
is in the context of expansions, the Kronecker delta remains as an
indication of the constraints between the indexes $\nu _{1}$ and $\nu _{2}$.

\subsubsection{Property ll. Multiregion Expansion (MRE)}

\qquad Consider the binomial expression:

\begin{equation}
\left( A_{1}+A_{2}\right) ^{\pm \nu },  \label{f13}
\end{equation}%
where the quantities $A_{1}$,$A_{2}$ and $\nu $, can take arbitrary values.
In this case there are two limits or possible regions with respect to the
expansion: the region where $\left( A_{1}>A_{2}\right) $ and the region
where $\left( A_{1}<A_{2}\right) $. These expansions are respectively:

\begin{enumerate}
\item Region $\left( A_{1}>A_{2}\right) $
\end{enumerate}

\begin{equation}
\left( A_{1}+A_{2}\right) ^{\pm \nu }=A_{1}^{\pm \nu
}\dsum\limits_{n=0}^{\infty }\dfrac{\left( \mp \nu \right) _{n}}{\Gamma (n+1)%
}\left( -\dfrac{A_{2}}{A_{1}}\right) ^{n}.  \label{f15}
\end{equation}

\begin{enumerate}
\item[2.] Region $\left( A_{1}<A_{2}\right) $
\end{enumerate}

\begin{equation}
\left( A_{1}+A_{2}\right) ^{\pm \nu }=A_{2}^{\pm \nu
}\dsum\limits_{n=0}^{\infty }\dfrac{\left( \mp \nu \right) _{n}}{\Gamma (n+1)%
}\left( -\dfrac{A_{1}}{A_{2}}\right) ^{n}.  \label{f14}
\end{equation}%
%
%
%
%
%
%
%
%
%
%
%
%
%
%
%
%
%
%
%
%
%
%
%
%
%
%
%
%
%
%
%
%
%
%
%
%
%
%
%
%
%
%
%
%
%
%
%
%
%
%
%
%
%
%
%
%
%
%
%
%
%
%
%
%
%
%
%
%
%
%
%
%
%
%
%
%
%
%
%
%
%
%
%
%
%
%
The factor $(\nu )_{n}$ is called Pochhammer symbol and is defined as:

\begin{equation}
(\nu )_{n}=\dfrac{\Gamma \left( \nu +n\right) }{\Gamma \left( \nu \right) }.
\end{equation}%
We have thus obtained separate expansions in the two possible limits.
Nevertheless, it is possible to express both results using a single series
that contains both limiting regions. In this sense we can say that this type
of expansion corresponds to a multiregion series representation of the
binomial $\left( \ref{f13}\right) $, through the use of the integral
representation of the denominator indicated in $\left( \ref{f1}\right) $.
Then we have:

\begin{equation}
\left( A_{1}+A_{2}\right) ^{\pm \nu }=\frac{1}{\Gamma (\mp \nu )}%
\int\limits_{0}^{\infty }dx\;x^{\mp \nu -1}\exp (-xA_{1})\exp (-xA_{2}),
\end{equation}%
and the exponentials are expanded separately, with the result:

\begin{equation}
\left( A_{1}+A_{2}\right) ^{\pm \nu }=\frac{1}{\Gamma (\mp \nu )}%
\dsum\limits_{n_{1}}\dsum\limits_{n_{2}}\frac{(-1)^{n_{1}+n_{2}}}{\Gamma
(n_{1}+1)\Gamma (n_{2}+1)}A_{1}^{n_{1}}A_{2}^{n_{2}}\int dx\;x^{\mp \nu
+n_{1}+n_{2}-1}.
\end{equation}%
Making use of the identity $\left( \ref{f4}\right) $ we get the multiregion
binomial expansion:

\begin{equation}
\left( A_{1}+A_{2}\right) ^{\pm \nu }=\frac{1}{\Gamma (\mp \nu )}%
\dsum\limits_{n_{1}}\dsum\limits_{n_{2}}\dfrac{(-1)^{n_{1}+n_{2}}}{\Gamma
(n_{1}+1)\Gamma (n_{2}+1)}A_{1}^{n_{1}}A_{2}^{n_{2}}\left\langle \mp \nu
+n_{1}+n_{2}\right\rangle ,  \label{f8}
\end{equation}%
where the parenthesis $\left\langle \cdot \right\rangle $, according to
property $\left( \ref{f10}\right) $, can be expressed this time in three
different ways, although in each one of them we will have the same Kronecker
delta which eliminates one of the two sums that are present:%
\begin{equation}
\begin{array}{cc}
\left\langle \mp \nu +n_{1}+n_{2}\right\rangle = & \left\{
\begin{array}{l}
\Gamma \left( \mp \nu +n_{1}\right) \dfrac{\Gamma \left( n_{2}+1\right) }{%
\left( -1\right) ^{n_{2}}}\;\delta _{\mp \nu +n_{1}+n_{2},0}, \\
\\
\Gamma \left( \mp \nu +n_{2}\right) \dfrac{\Gamma \left( n_{1}+1\right) }{%
\left( -1\right) ^{n_{1}}}\;\delta _{\mp \nu +n_{1}+n_{2},0}, \\
\\
\Gamma \left( n_{2}+n_{1}\right) \dfrac{\Gamma \left( \mp \nu +1\right) }{%
\left( -1\right) ^{\mp \nu }}\;\delta _{\mp \nu +n_{1}+n_{2},0}.%
\end{array}%
\right.%
\end{array}%
\end{equation}%
On the other hand, the number of possible forms of summing $\left( \ref{f8}%
\right) $ using for this purpose the Kronecker delta, can be found
evaluating the combinatorics $C_{Deltas}^{Sums}$, which in this case is $%
C_{1}^{2}=2$. Let us see what happens if one sums with respect to a
particular index:

\begin{enumerate}
\item \textbf{Sum respect to} $n_{2}$
\end{enumerate}

Using for this case the following equality:

\begin{equation}
\left\langle \mp \nu +n_{1}+n_{2}\right\rangle =\Gamma \left( \mp \nu
+n_{1}\right) \dfrac{\Gamma \left( n_{2}+1\right) }{\left( -1\right) ^{n_{2}}%
}\;\delta _{\mp \nu +n_{1}+n_{2},0},
\end{equation}%
and then replacing in $\left( \ref{f8}\right) $, we get:

\begin{equation}
\left( A_{1}+A_{2}\right) ^{\pm \nu }=\frac{1}{\Gamma (\mp \nu )}%
\dsum\limits_{n_{1}}(-1)^{n_{1}}\dfrac{\Gamma \left( \mp \nu +n_{1}\right) }{%
\Gamma (n_{1}+1)}A_{1}^{n_{1}}A_{2}^{\pm \nu -n_{1}},
\end{equation}%
or equivalently:

\begin{equation}
\left( A_{1}+A_{2}\right) ^{\pm \nu }=A_{2}^{\pm \nu
}\dsum\limits_{n_{1}=0}^{\infty }\dfrac{\left( \mp \nu \right) _{n_{1}}}{%
\Gamma (n_{1}+1)}\left( -\frac{A_{1}}{A_{2}}\right) ^{n_{1}},
\end{equation}%
which gives the expression associated to the region $\left(
A_{1}<A_{2}\right) $, obtained previously in $\left( \ref{f14}\right) $.

\begin{enumerate}
\item[2.] \textbf{Sum respect to} $n_{1}$
\end{enumerate}

Analogously, we now use the identity:

\begin{equation}
\left\langle \mp \nu +n_{1}+n_{2}\right\rangle =\Gamma \left( \mp \nu
+n_{2}\right) \dfrac{\Gamma (n_{1}+1)}{\left( -1\right) ^{n_{1}}}\;\delta
_{\mp \nu +n_{1}+n_{2},0},
\end{equation}%
and replacing in $\left( \ref{f8}\right) $, one gets:

\begin{equation}
\left( A_{1}+A_{2}\right) ^{\pm \nu }=A_{1}^{\pm \nu
}\dsum\limits_{n_{2}=0}^{\infty }\frac{\left( \mp \nu \right) _{n_{2}}}{%
\Gamma (n_{2}+1)}\left( -\frac{A_{2}}{A_{1}}\right) ^{n_{2}},
\end{equation}%
an expression that was already found in $\left( \ref{f15}\right) $, and
valid in the region $\left( A_{1}>A_{2}\right) $.

The fundamental idea that has been exposed in the previous demonstration is
that using the definition $\left( \ref{f2}\right) $ one can do a binomial
expansion that differs from the conventional expansion, in the sense that
one obtains an expression where all the possible limits of the binomial are
included simultaneously.

\subsection{Multiregion expansion of a multinomial}

\qquad The previous discussion can be generalized to the multiregion
expansion of a multinomial of $n_{l}$ terms:

\begin{equation}
\left( A_{1}+...+A_{l}\right) ^{\pm \nu
}=\dsum\limits_{n_{1}}...\dsum\limits_{n_{l}}\phi
_{n_{1},..,n_{l}}\;A_{1}^{n_{1}}...A_{l}^{n_{l}}\frac{\left\langle \mp \nu
+n_{1}+...+n_{l}\right\rangle }{\Gamma (\mp \nu )},  \label{f9}
\end{equation}%
where for simplicity we have defined the following notation:

\begin{equation}
\phi _{n_{1},..,n_{l}}=(-1)^{n_{1}+...+n_{l}}\;\frac{1}{\Gamma
(n_{1}+1)..\Gamma (n_{l}+1)}.
\end{equation}%
The number of expansions that can be obtained starting from $\left( \ref{f9}%
\right) $ is given at most by all the possible forms of evaluating some of
the sums, using for this the Kronecker delta generated by the same
expansion, that is $C_{1}^{n_{l}}=n_{l}$ possible forms. More generally, a
function expressed as a multiregion expansion through $\sigma $ sums and $%
\delta $ Kronecker deltas, can be evaluated at most in:

\begin{equation}
C_{\delta }^{\sigma }=\dfrac{\sigma !}{\delta !(\sigma -\delta )!}
\end{equation}%
possible forms, each one of them expressed in terms of series of
multiplicity $\mu =\left( \sigma -\delta \right) $. All the resulting series
are representations with respect to the ratios between the terms of the
multinomial, and all of them correspond to multivariable generalizations of
the hypergeometric function.

\section{Negative dimension $D$ ?}

\qquad The original name of the integration method here presented, NDIM
(Negative Dimension Integration Method), comes from applying the expansion
and subsequent association, integral$\Leftrightarrow $Kronecker delta, over
the gaussian integral in $D$ dimensions:

\begin{equation}
\dint \frac{d^{D}k}{i\pi ^{\frac{D}{2}}}\;\exp (\alpha k^{2}),  \label{f36}
\end{equation}%
where $k$ corresponds to a 4-momentum and $\alpha $ is an arbitrary
parameter. This is a typical integral that appears upon using Schwinger's
parametrization and after using the completion of squares procedure of the
loop momenta. Starting from this equation it is possible to deduce in a
similar way to equation $\left( \ref{f28}\right) $ the following identity
\cite{IRi}:

\begin{equation}
\dint \frac{d^{D}k}{i\pi ^{\frac{D}{2}}}\;(k^{2})^{n}=n!\;\delta _{n+\frac{D%
}{2},0}.  \label{f29}
\end{equation}%
Since this is in fact a Taylor expansion, we assume that $n\geq 0$, and
therefore the constraint associated to the Kronecker delta in $\left( \ref%
{f29}\right) $\ requires that $\frac{D}{2}\leq 0$, that is, la dimension has
to be negative, which is then the origin of the name of this integration
technique. Nevertheless, there can appear reasonable doubts with respect to
some concepts that the NDIM uses in its original approach.

It is entirely possible for $D$ to be a negative integer considering that
the Feynman integrals are analytic functions in $D$ arbitrary dimensions,
but rigorously what is wanted is to solve the loop integrals in the limit $%
D\rightarrow +4$ and not in the limit $D\rightarrow -4$. The inconsistency
is even more clear if the dimensional regularization prescription is used,
and the dimension $D$ now includes a non integer piece, the dimensional
regulator $\epsilon $, which cannot be associated to the sum index $n$ in $%
\left( \ref{f29}\right) $.

Such inconsistencies could be resolved if the expansion of the exponential
could be written as:

\begin{equation}
\begin{array}{ccc}
\exp (x)=\lim\limits_{\xi \rightarrow 0}\dsum\limits_{m=-\infty +\xi
}^{\infty +\xi }\dfrac{x^{m}}{\Gamma (m+1)} &  & \text{such that }\left.
\dfrac{d^{m}\exp (x)}{dx^{m}}\right\vert _{x=0}=1,%
\end{array}
\label{f45}
\end{equation}%
where $\left\vert \xi \right\vert <1$ and $m$ is an index that increases in
one unit steps. Such an expansion is possible, but for this purpose it is
necessary to resort to the area of calculus called Fractional Calculus \cite%
{RHi, KMi, KOl}, through which it can be shown that both the derivative and
the integration operators can be represented in terms of a unique operator,
and in second term shows that the analytical continuation in the operator
order can in fact be done, and can be fractional and even complex. This is
very useful in order to justify the validity of the NDIM integration
technique, and therefore it is necessary to review some concepts for
understanding this type of calculus.

\subsection{Preliminaries : The fractional expansion of the exponential
function.}

\qquad We are interested in knowing the nature of the expansion that is
performed in the exponential function in $\left( \ref{f36}\right) $, since
due to the arguments mentioned before it should not really correspond to a
Taylor expansion.

In order to provide a more rigorous basis for equation $\left( \ref{f45}%
\right) $, it becomes necessary to define the integration and derivative
operators in a generalized form. In fractional calculus is possible to
define the fractional integral of order $\alpha $ as follows:

\begin{equation}
\;_{c}D_{x}^{-\alpha }f(x)=\frac{1}{\Gamma (\alpha )}\int\limits_{c}^{x}dt\;
\frac{f(t)}{(x-t)^{1-\alpha }},  \label{f30}
\end{equation}%
where $c\in
\mathbb{R}
$ and $\alpha $ is an arbitrary quantity. For the particular case in which $%
c=0 $ in $\left( \ref{f30}\right) $, we get the so called Riemann-Liouville
fractional integral:

\begin{equation}
\;_{0}D_{x}^{-\alpha }f(x)=\frac{1}{\Gamma (\alpha )}\int\limits_{0}^{x}dt\;%
\frac{f(t)}{(x-t)^{1-\alpha }}.
\end{equation}%
Another version of the fractional integral can be obtained making $c=-\infty
$, which is called Liouville fractional integral:

\begin{equation}
\;_{-\infty }D_{x}^{-\alpha }f(x)=\frac{1}{\Gamma (\alpha )}%
\int\limits_{-\infty }^{x}dt\;\frac{f(t)}{(x-t)^{1-\alpha }}.  \label{f32}
\end{equation}%
This last version of the fractional integral will be used in this work in
order to justify more rigorously the integration method of Feynman diagrams
here presented. Certainly this would be incomplete if we do not define also
the fractional derivative (in the Liouville version). For an order $\alpha $
which fulfills that $(m-1)<\alpha \leq m$, with $m\in
\mathbb{N}
$, the fractional derivative is defined by the expression $_{-\infty
}D_{x}^{\alpha }f(x)=\dfrac{d^{m}}{dx^{m}}\;\left[ _{-\infty
}D_{x}^{m-\alpha }f(x)\right] $, or in operational terms:

\begin{equation}
_{-\infty }D_{x}^{\alpha }f(x)=\left\{
\begin{array}{ll}
\dfrac{1}{\Gamma (m-\alpha )}\dfrac{d^{m}}{dx^{m}}\dint\limits_{-\infty
}^{x}dt\;\dfrac{f(t)}{(x-t)^{1+\alpha -m}} & ,\;(m-1)<\alpha <m, \\
&  \\
\dfrac{d^{m}}{dx^{m}}f(x) & ,\;\alpha =m.%
\end{array}%
\right.  \label{f33}
\end{equation}%
We are particularly interested in the Liouville integro-differential
operator, so for simplifying the notation let us define $_{-\infty
}D_{x}^{\alpha }=D_{x}^{\alpha }$, where we can have $\left( \alpha
<0\right) $, which is an integral, or $\left( \alpha >0\right) $, a
derivative.

A property that makes this version of the fractional operator particularly
interesting is the effect that it has over the exponential function, and
which in practice is an extension of what is done by integrals and
derivatives of integer order (conventional calculus) over this function:

\begin{equation}
D_{x}^{\alpha }\left[ \exp (\beta x)\right] =\left\{
\begin{array}{l}
\beta ^{-\alpha }\exp (\beta x) \\
\\
\beta ^{\alpha }\exp (\beta x)%
\end{array}%
\right.
\begin{array}{l}
\alpha <0\;\text{(integration of order }\alpha \text{),} \\
\\
\alpha >0\;\text{(derivation of order }\alpha \text{).}%
\end{array}
\label{f44}
\end{equation}%
This expression can be easily proved using the equations $\left( \ref{f32}%
\right) $ and $\left( \ref{f33}\right) $.

\subsection{Taylor and Taylor-Riemann series. The fractional expansion of
diagram $G$}

\qquad Particular interest has the series representation of the exponential
function, since it is the only function over which the expansions in the
Schwinger parameter integral are done, either because it appears explicitly
in the integrand, or because it is used implicitly in order to do the
multiregion expansions of the multinomials present in the problem. When the
expansions in the integral of the parametric integral are done, we
automatically assume that this is a Taylor series. Nevertheless, what is
really done implicitly is a Taylor-Riemann expansion of the exponential,
which reads:

\begin{equation}
\exp (x)=\dsum\limits_{n=-\infty }^{\infty }\left. D_{x}^{n+\xi }\left[ \exp
(x)\right] \right\vert _{x=0}\dfrac{x^{n+\xi }}{\Gamma (n+\xi +1)}.
\end{equation}%
The parameter $\ \xi $ is arbitrary and fulfills the condition $0\leq
\left\vert \xi \right\vert <1$. Notice that taking $\xi \rightarrow 0$ the
Taylor series for the exponential is obtained. Moreover, since $\left.
D_{x}^{n+\xi }\left[ \exp (x)\right] \right\vert _{x=0}=1$, according to
formula $\left( \ref{f44}\right) $, one gets:

\begin{equation}
\exp (x)=\dsum\limits_{n=-\infty }^{\infty }\dfrac{x^{n+\xi }}{\Gamma (n+\xi
+1)}.
\end{equation}%
If we make now the change of variables $m=n+\xi $, we can rewrite the
previous expansion as:

\begin{equation}
\exp (x)=\sum\limits_{m=-\infty +\xi }^{\infty +\xi }\dfrac{x^{m}}{\Gamma
(m+1)},  \label{f31}
\end{equation}%
which looks like the conventional Taylor series expansion of the
exponential, with the difference that the index of the sum has been
analytically continued to negative and fractional values. This justifies
that both $D$ and the propagator powers $\left\{ \nu _{1},...,\nu
_{N}\right\} $ do not change its nature due to some analytical continuation
required by the Kronecker deltas generated in the process, but rather are
the expansion indexes the ones that can acquire arbitrary values, including
negative and fractional. But moreover it justifies the presence of Gamma
functions in the multiregion expansion of the diagram, both in the numerator
and denominator, which having the form $\Gamma (-n)$, where $n$ is an index
associated to a sum, give a multiregion expansion not necessarily infinite
or vanishing.

After having said that, it is still necessary to justify what refers to the
integration method, since they correspond to hypergeometric series whose
summation indexes are positive integers. This can be explained by saying
that for applying the integration method NDIM an analytical continuation of
the summation indexes has to be done, or equivalently the exponentials are
expanded in Taylor-Riemann series, using the expression $\left( \ref{f31}%
\right) $. In practical terms it is enough to assume a fractional arbitrary
parameter $\xi $ associated to the indexes of all the sums, and at the end
of the integration process, when each contribution to the solution is
determined, evaluate the limit:

\begin{equation}
\lim_{\xi \rightarrow 0}\dsum\limits_{m=-\infty +\xi }^{\infty +\xi
}\Longrightarrow \dsum\limits_{m=0}^{\infty }\;,
\end{equation}%
which naturally happens since all the solutions that are obtained have the
factor $\dfrac{1}{\Gamma (m+1+\xi )}$, and in the limit $\xi \rightarrow 0$
we have that:

\begin{equation}
\lim_{\xi \rightarrow 0}\dfrac{1}{\Gamma (m+1+\xi )}=\left\{
\begin{array}{l}
\dfrac{1}{m!} \\
\\
0%
\end{array}%
\right.
\begin{array}{l}
\text{, para }m\geq 0, \\
\\
\text{, para }m<0.%
\end{array}%
\end{equation}%
Therefore, strictly speaking we should write the Multiregion Expansion $%
\left( \ref{f23}\right) $ of a generic diagram $G$ in the following form:

\begin{equation}
\begin{array}{ll}
G= & (-1)^{-\frac{LD}{2}}\lim_{\xi _{1},..,\xi _{\sigma }\rightarrow
0}\dsum\limits_{n_{1}=-\infty +\xi _{1}}^{\infty +\xi
_{1}}...\dsum\limits_{n_{\sigma }=-\infty +\xi _{\sigma }}^{\infty +\xi
_{\sigma }}\phi _{n_{1},..,n_{\sigma
}}\;\tprod\limits_{j=1}^{P}(Q_{j}^{2})^{n_{j}}\tprod%
\limits_{j=P+1}^{P+M}(-m_{j}^{2})^{n_{j}} \\
&  \\
& \tprod\limits_{j=1}^{N}\dfrac{\left\langle \nu _{j}+\alpha
_{j}\right\rangle }{\Gamma (\nu _{j})}\tprod\limits_{j=1}^{K}\dfrac{%
\left\langle \beta _{j}+\gamma _{j}\right\rangle }{\Gamma (\beta _{j})}.%
\end{array}%
\end{equation}%
Operationally NDIM does not consider the presence of the fractional part $%
\xi $ in the summation index nor its extension to negative values.
Nevertheless, the analytic continuation of the summation indexes and the
evaluation of the limit at the end of the process, are inherent to the
process and justify why it works. A confusion might arise from the
similarity of the Taylor series expansion of the exponential and its
Taylor-Riemann expansion (equation $\left( \ref{f31}\right) $), something
that explains why both versions of the expansion give the same result, which
in turn shows that there is a conceptual but not operational difference.
Thus it would be more appropriate to call the method integration by
fractional expansion (IBFE) instead of NDIM.

\section{Hypergeometric Functions of one and two variables}

\qquad In this work we have developed the evaluation of loop integrals of
certain class of diagrams with arbitrary $L$. All the found solutions have
been presented in terms of hypergeometric series of one or two variables,
which correspond in a natural way to an serie representation with respect to
the ratio between the energy scales associated to the graph. The purpose of
this appendix is precisely to provide the necessary information about the
hypergeometric functions \cite{WBa, GGa, LSl, LGr} and especially with
respect to their convergence properties.

\subsection{Definition of the generalized hypergeometric function}

\qquad In those cases in which the solutions contain two energy scales the
series representations are expressed in terms of generalized hypergeometric
functions:

\begin{equation}
\;_{q}F_{q-1}(a_{1},...,a_{q};b_{1},...,b_{q-1};z)\equiv \;_{q}F_{q-1}\left(
\left.
\begin{array}{c}
\left\{ a\right\} \\
\left\{ b\right\}%
\end{array}%
\right\vert z\right) =\dsum\limits_{k=0}^{\infty }\frac{\left( a_{1}\right)
_{k}...\left( a_{q}\right) _{k}}{\left( b_{1}\right) _{k}...\left(
b_{q-1}\right) _{k}}\frac{z^{k}}{k!},
\end{equation}%
where the factors $\left( \alpha \right) _{k}$ are called Pochhammer
symbols, defined as:

\begin{equation}
\left( \alpha \right) _{k}=\dfrac{\Gamma (\alpha +k)}{\Gamma (\alpha )}.
\end{equation}%
The convergence conditions for these series are given according to the
magnitude of the argument:

\begin{enumerate}
\item If $\left\vert z\right\vert <1$ the series converges absolutely. The
variable $z$ represents the ratio between energy scales of the topology.

\item If $z=1$, the necessary requirement for the convergence of the series
is that $\func{Re}\left( \omega \right) >0$, where $\omega $ is called
parametric excess and is given by:
\end{enumerate}

\begin{equation}
\begin{array}{l}
\omega =\dsum\limits_{j=0}^{q}b_{j}-\dsum\limits_{j=0}^{q+1}a_{j}.%
\end{array}%
\end{equation}%
For the convergence when $z=-1$ it is sufficient that $\func{Re}\left(
\omega \right) >-1$.

\subsection{Some identities of the Pochhammer symbols}

\qquad The following identities are useful when building the hypergeometric
function from the contributions that in turn are obtained from the
multiregion expansion of an arbitrary diagram $G$. The Pochhammer symbols
are defined as follows:

\begin{equation}
\left( a\right) _{n}=\left\{
\begin{array}{lll}
\tprod\limits_{j=0}^{n-1}(a+j) &  & \text{si }n>0, \\
&  &  \\
1 &  & \text{si }n=0,%
\end{array}%
\right.
\end{equation}%
Nevertheless, the series that are the result of applying the NDIM to Feynman
integrals will always contain factors of the type $\Gamma (a\pm n)$ and $%
\Gamma (a\pm 2n)$. In these cases it is convenient to use the following
formulae for finding the hypergeometric representations:

\begin{equation}
\left( a\right) _{n}=\dfrac{\Gamma (a+n)}{\Gamma (a)},
\end{equation}

\begin{equation}
\left( a\right) _{-n}=\dfrac{\Gamma (a-n)}{\Gamma (a)}=\dfrac{(-1)^{n}}{%
(1-a)_{n}},
\end{equation}

\begin{equation}
(a)_{2n}=\dfrac{\Gamma (a+2n)}{\Gamma (a)}=4^{n}\left( \dfrac{a}{2}\right)
_{n}\left( \dfrac{a}{2}+\dfrac{1}{2}\right) _{n}.
\end{equation}

\subsection{Two-variable hypergeometric functions}

\qquad In this part of the appendix we describe the functions which are
useful for representing those cases that consider three different energy
scales, in the general case $L>1$. Such functions correspond to the de Kamp%
\'{e} de F\'{e}riet generalized double hypergeometric function $F^{\substack{
p:r:u  \\ q:s:v}}$ and the generalized hypergeometric $\overline{F}
^{\substack{ p:r:u  \\ q:s:v}}$. The first is defined as:

\begin{equation}
\begin{array}{ll}
F^{\substack{ p:r:u  \\ q:s:v}}\left( \left.
\begin{array}{ccc}
\alpha _{1},..,\alpha _{p} & a_{1},..,a_{r} & c_{1},..,c_{u} \\
\beta _{1},..,\beta _{q} & b_{1},..,b_{s} & d_{1},..,d_{v}%
\end{array}%
\right\vert x,y\right) & =F^{\substack{ p:r:u  \\ q:s:v}}\left( \left.
\begin{array}{ccc}
\{\alpha \} & \{a\} & \{c\} \\
\{\beta \} & \{b\} & \{d\}%
\end{array}%
\right\vert x,y\right) \\
&  \\
& =\dsum\limits_{n,m}^{\infty }\dfrac{\tprod\limits_{j=1}^{p}(\alpha
_{j})_{n+m}\tprod\limits_{j=1}^{r}(a_{j})_{n}\tprod%
\limits_{j=1}^{u}(c_{j})_{m}}{\tprod\limits_{j=1}^{q}(\beta
_{j})_{n+m}\tprod\limits_{j=1}^{s}(b_{j})_{n}\tprod%
\limits_{j=1}^{v}(d_{j})_{m}}\dfrac{x^{n}}{n!}\dfrac{y^{m}}{m!},%
\end{array}%
\end{equation}%
where the convergence of the double series requires that the following
relation between the indexes is fulfilled:

\begin{equation}
p+r\leqslant q+s+1,
\end{equation}

\begin{equation}
p+u\leqslant q+v+1,
\end{equation}%
and also that the arguments fulfill the condition:

\begin{equation*}
\begin{array}{lll}
\left\vert x\right\vert ^{\tfrac{1}{(p-q)}}+\left\vert y\right\vert ^{\tfrac{%
1}{(p-q)}}<1 &  & \text{, for the case in which }(p>q), \\
&  &  \\
\max \{\left\vert x\right\vert ,\left\vert y\right\vert \}<1 &  & \text{, in
the case }(p\leqslant q).%
\end{array}%
\end{equation*}%
For the other series that occurs frequently we have the following definition:

\begin{equation}
\begin{array}{ll}
\overline{F}^{\substack{ p:r:u  \\ q:s:v}}\left( \left.
\begin{array}{ccc}
\alpha _{1},..,\alpha _{p} & a_{1},..,a_{r} & c_{1},..,c_{u} \\
\beta _{1},..,\beta _{q} & b_{1},..,b_{s} & d_{1},..,d_{v}%
\end{array}%
\right\vert x,y\right) & =\overline{F}^{\substack{ p:r:u  \\ q:s:v}}\left(
\left.
\begin{array}{ccc}
\{\alpha \} & \{a\} & \{c\} \\
\{\beta \} & \{b\} & \{d\}%
\end{array}%
\right\vert x,y\right) \\
&  \\
& =\dsum\limits_{n,m}^{\infty }\dfrac{\tprod\limits_{j=1}^{p}(\alpha
_{j})_{n-m}\tprod\limits_{j=1}^{r}(a_{j})_{n}\tprod%
\limits_{j=1}^{u}(c_{j})_{m}}{\tprod\limits_{j=1}^{q}(\beta
_{j})_{n-m}\tprod\limits_{j=1}^{s}(b_{j})_{n}\tprod%
\limits_{j=1}^{v}(d_{j})_{m}}\dfrac{x^{n}}{n!}\dfrac{y^{m}}{m!},%
\end{array}%
\end{equation}%
where the convergence of the double series requires that the following
relations between the indexes are satisfied:

\begin{equation}
p+r\leqslant q+s+1,
\end{equation}

\begin{equation}
q+u\leqslant p+v+1.
\end{equation}%
All solutions obtained with the technique of integration NDIM fullfill the
previous condition in the sum indexes. The conditions of convergence of
arguments may be worked out using Horns general theory of convergence \cite%
{HEx}.

\newpage

\section{Bibliograf\'{\i}a}


\begin{thebibliography}{99}
\bibitem{VSm} V.A.Smirnov, Evaluating Feynman Integrals (Springer, Berlin,
Heidelberg, 2004); and references therein.

\bibitem{SMo} S.Moch, P.Uwer, S.Weinzierl, J.Math.Phys. 43 (2002) 3363,
(hep-ph/0110083).

\bibitem{DMa} T.Huber, D.Maitre, Comput.Phys.Commun. 175 (2006) 122-144,
(hep-ph/0507094).

\bibitem{MKa} M.Yu.Kalmykov JHEP 0604 (2006) 056, (hep-th/0602028).

\bibitem{IRi} I.G.Halliday, R.M. Ricotta, Phys. Lett. B 193\textbf{\ }(1987)
241.

\bibitem{CAn2} C.Anastasiou, E.W.N.Glover, C.Oleari, Nucl. Phys. B 572
(2000) 307, (hep-ph/9907494).

\bibitem{ASu11} A.T.Suzuki, E.S.Santos, A.G.M.Schmidt, Eur.Phys.J. C 26
(2002) 125, (hep-th/0205158).

\bibitem{ASu13} A.T.Suzuki, E.S.Santos, A.G.M.Schmidt, J.Phys. A36 (2003)
4465, (hep-ph/0210148).

\bibitem{ASu14} A.T.Suzuki, A.G.M.Schmidt, J.Phys. A31 (1998) 8023.

\bibitem{CAn1} C.Anastasiou, E.W.N.Glover, C.Oleari, Nucl.Phys. B 565 (2000)
445, (hep-ph/9907523).

\bibitem{COl} C.Anastasiou, E.W.N.Glover, C.Oleari,Nucl.Phys. B 575 (2000)
416. [Erratum-ibid. B 585 (2000) 763].

\bibitem{ASu2} A.T.Suzuki, A.G.M.Schmidt, Can.J.Phys. 78 (2000) 769,
(hep-th/9904195).

\bibitem{ASu3} A.T.Suzuki, A.G.M.Schmidt, JHEP 9709 (1997) 002,
(hep-th/9709024).

\bibitem{ASu4} A.T.Suzuki, A.G.M.Schmidt, Eur.Phys.J. C 5 (1998) 175,
(hep-th/9709144).

\bibitem{ASu5} A.T.Suzuki, A.G.M.Schmidt, Solutions for a massless off-shell
two-loop three-point vertex, (hep-th/9712104).

\bibitem{ASu6} A.T.Suzuki, A.G.M. Schmidt, Phys.Rev. D58 (1998) 047701,
(hep-th/9712108).

\bibitem{ASu10} A.T.Suzuki, A.G.M.Schmidt, J.Phys. A35 (2002) 151,
(hep-th/0110047).

\bibitem{IGo} I.Gonzalez, I.Schmidt, Phys.Rev. D72 (2005) 106006,
(hep-th/0508013).

\bibitem{EBo} E.E.Boos, A.I.Davydychev, Theor. Math. Phys. 89 (1991) 1052.

\bibitem{ADa1} A.I.Davydychev, J. Math.Phys.32 (1991) 1052.

\bibitem{ADa2} A.I.Davydychev, J. Math.Phys.33 (1992) 358.

\bibitem{ASu15} A.T.Suzuki, A.G.M. Schmidt, J.Comput.Phys. 168 (2001) 207,
(hep-th/0008143).

\bibitem{RHi} R.Hilfer, Applications of Fractional Calculus in Physics
(World Scientific Publishing Co. Pte. Ltd., Singapore, 2000).

\bibitem{KMi} K.S.Miller, B.Ross, An introduction to the fractional calculus
and fractional differential equations (John Wiley \& Sons, Inc., 1993).

\bibitem{KOl} K.B.Oldham, J.Spanier, The Fractional Calculus\textit{,}
Theory and Applications of Differentiation and Integration to Arbitrary
Order (Academic Press, Inc., 1974).

\bibitem{WBa} W.N.Bailey, Generalized Hypergeometric Functions (Cambridge
University Press, Cambridge, 1966).

\bibitem{GGa} G.Gasper, M.Rahman, Basic Hypergeometric Series (Cambridge
University Press, 1990).

\bibitem{LSl} L.J.Slater, Generalized Hypergeometric Functions (Cambridge
University Press, 1966).

\bibitem{LGr} L.S.Gradshteyn, L.M.Ryzhik, Table of Integrals, Series, and
Products, Sixth Edition (Academic Press, 2000).

\bibitem{HEx} H. Exton, \textit{Multiple Hypergeometric Functions and
Applications} (Ellis Horwood,Westergate, England, 1976).
\end{thebibliography}
\end{document}